\documentclass[journal,10pt,twocolumn]{IEEEtran}
\usepackage{graphicx}
\usepackage{subcaption}
\usepackage{amsmath,amssymb,amsfonts}
\usepackage{color}
\usepackage{multirow}
\usepackage{textcomp}
\usepackage{ragged2e}
\usepackage{epstopdf}
\usepackage{url}
\usepackage{comment}
\usepackage[utf8]{inputenc}
\usepackage[T1]{fontenc}

\ifCLASSOPTIONcompsoc
  \usepackage[nocompress]{cite}
\else
  \usepackage{cite}
\fi
\begin{document}
\title{Qutrit-inspired Fully Self-supervised Shallow Quantum Learning Network for Brain Tumor Segmentation}

\author{Debanjan~Konar,~\IEEEmembership{MIEEE,}
        Siddhartha~Bhattacharyya,~\IEEEmembership{SMIEEE,}
        Bijaya~K.~Panigrahi,~\IEEEmembership{SMIEEE,} and 
         Elizabeth~Behrman
\thanks{D.~Konar and B.~K.~Panigrahi are with the Department of Electrical Engineering, Indian Institute of Technology Delhi, New Delhi, India, Email: konar.debanjan@ieee.org and bkpanigrahi@ee.iitd.ernet.in}
\thanks{S.~Bhattacharyya is with the Department of Computer Science and Engineering, CHRIST (Deemed to be University), Bangalore, India, Email: dr.siddhartha.bhattacharyya@gmail.com India}
\thanks{E.~Behrman is with the Department of Mathematics and Physics, Wichita State University, Wichita, Kansas, USA, Email: behrman@math.wichita.edu}}



\maketitle

\begin{abstract}
 Classical self-supervised networks suffer from convergence problems and reduced segmentation accuracy due to forceful termination. \emph{Qubits} or bi-level quantum bits often describe quantum neural network models. In this article, a novel self-supervised shallow learning network model exploiting the sophisticated three-level qutrit-inspired quantum information system referred to as Quantum Fully Self-Supervised Neural Network (QFS-Net) is presented for automated segmentation of brain MR images. The QFS-Net model comprises a trinity of a layered structure of \emph{qutrits} inter-connected through parametric Hadamard gates using an 8-connected second-order neighborhood-based topology. The non-linear transformation of the \emph{qutrit} states allows the underlying quantum neural network model to encode the quantum states, thereby enabling a faster self-organized counter-propagation of these states between the layers without supervision. The suggested QFS-Net model is tailored and extensively validated on Cancer Imaging Archive (TCIA) data set collected from Nature repository and also compared with state of the art supervised (U-Net and URes-Net architectures) and the self-supervised QIS-Net model. Results shed promising segmented outcome in detecting tumors in terms of dice similarity and accuracy with minimum human intervention and computational resources.
\end{abstract}

\begin{IEEEkeywords}
Quantum Computing, Qutrit, QIS-Net, MR image segmentation, U-Net and URes-Net.
\end{IEEEkeywords}

 \ifCLASSOPTIONpeerreview
\begin{center} \bfseries Quantum Inspired Self-Supervised Neural Network Architecture \end{center}
\fi
%
\IEEEpeerreviewmaketitle

\section{Introduction}
\label{chapter:intro}

\IEEEPARstart{Q}{}uantum computing supremacy may be achieved through the superposition of quantum states or quantum parallelism and quantum entanglement~\cite{oster}. However, owing to the lack of computing resources for the implementation of quantum algorithms, it is an uphill task to explore the quantum entanglement properties for optimized computation. Nowadays, with the advancement in quantum algorithms, the classical systems embedded in quantum formalism and inspired by \emph{qubits} cannot exploit the full advantages of quantum superposition and quantum entanglement~\cite{gandhi, chen, xiao}. Due to the intrinsic characteristics offered by quantum mechanics, the implementation of Quantum-inspired Artificial Neural Networks (QANN) has been proven to be successful in solving specific computing tasks like image classification, pattern recognition~\cite{lu, bhatt1, konar1, konar3}. Nevertheless, quantum neural network models~\cite{schuld, kapoor} implemented on actual quantum processors are realized using a large number of quantum bits or \emph{qubits} as matrix representation and as well as linear operations on these vector matrices. However, owing to complex and time-intensive quantum back-propagation algorithms involved in the supervised quantum-inspired neural network (QINN) architectures~\cite{narayan,liu}, the computational complexity increases many-fold with an increase in the number of neurons and inter-layer interconnections. \\
Automatic segmentation of brain lesions from Magnetic Resonance Imaging (MRI) greatly facilitates brain tumor identification overcoming the manual laborious tasks of human experts or radiologists~\cite{clark}. It contrasts with the manual brain tumor diagnosis which suffers from significant variations in shape, size, orientation, intensity inhomogeneity, overlapping of gray-scales and inter-observer variability. Recent years have witnessed substantial attention in developing robust and efficient automated MR image segmentation procedures among the researchers of the computer vision community. \\
The current work focuses on a novel quantum fully self-supervised learning network (QFS-Net) characterized by \emph{qutrits} for fast and accurate segmentation of brain lesions. The primary aim of the suggested work is to enable the QFS-Net for faster convergence and making it suitable for fully automated brain lesion segmentation obviating any kind of training or supervision. The proposed quantum fully self-supervised neural network (QFS-Net) model relies on \emph{qutrits} or three-level quantum states to exploit the features of quantum correlation. To eliminate the complex quantum back-propagation algorithms used in the supervised QINN models, the QFS-Net resorts to a novel fully self-supervised \emph{qutrit} based counter propagation algorithm. This algorithm allows the propagation of quantum states between the network layers iteratively. The primary contributions of our manuscript are fourfold and are highlighted as follows: 
\begin{enumerate}
	\item Of late, the quantum neural network models and their implementation largely rely on \emph{qubits} and hence, we have proposed a novel \emph{qudit} embedded generic quantum neural network model applicable for any level of quantum states such as \emph{qubit, qutrit} etc. 
	\item An adaptive multi-class Quantum Sigmoid (\emph{QSig}) activation function embedded with quantum trit or \emph{qutrit} is incorporated to address the wide variation of gray scales in MR images. 
	\item The convergence analysis of the QFS-Net model is provided, and its super-linearity is also demonstrated experimentally. The proposed \emph{qutrit} based quantum neural network model tries to explore the superposition and entanglement properties of quantum computing in classical simulations resulting in faster convergence of the network architecture yielding optimal segmentation.
	\item The suggested QFS-Net model is validated extensively using Cancer Imaging Archive (TCIA) data set collected from Nature repository~\cite{data}. Experimental results show the efficacy of the proposed QFS-Net model in terms of dice similarity, thus promoting self-supervised procedures for medical image segmentation.
\end{enumerate}

The remaining sections of the article are organized as follows. Section~\ref{review} reviews various supervised artificial neural networks and deep neural network models useful for brain MR image segmentation. A brief introduction of \emph{qutrits} and generalized $D$-level quantum states (\emph{qudits}) is provided along with the preliminaries of quantum computing in Section~\ref{QC}. A novel quantum neural network model characterized by \emph{qudits} is illustrated in Section~\ref{QINN:Add}. A vivid description of the suggested QFS-Net and its operations characterized by \emph{qutrit} has been provided in Section~\ref{QFS-Net}. Results and discussions shed light on the experimental outcomes of the proposed neural network model in Section~\ref{results:disscuss}. Concluding remarks and future directions of research are confabulated in Section~\ref{discuss}.

\section{Literature Review}
\label{review}

Recent years have witnessed various machine learning classifiers~\cite{lee, zikic} and deep learning technologies~\cite{zikic1, olaf, guerrero, brebisson} for automated brain lesion segmentation for tumor detection. Examples include U-Net~\cite{olaf} and UResNet~\cite{guerrero}, which have achieved remarkable dice score in auto-segmentation of medical images. Of late, Pereira \emph{et al.}~\cite{pereira} suggested a modified Convolutional Neural Network (CNN) introducing small size kernels to obviate over-fitting. Moreover, CNN based architectures suffers due to lack of manually segmented or annotated MR images, intensive pre-processing and expert image analysts. In these circumstances, self-supervised or semi-supervised medical image segmentation is becoming popular in the computer vision research community. Wang \emph{et al.}~\cite{wang} contributed an interactive method using deep learning with image-specific tuning for medical image segmentation. Zhuang \emph{et al.}~\cite{zhuang} suggested a Rubik's cube recovery based self-supervised procedure for medical image segmentation. However, the interactive learning frameworks are not fully self-supervised and suffer from the complex orientation and time-intensive operations.\\  
Quantum Artificial Neural Networks (QANN) were first proposed in the 1990s \cite{behrman, kak, ventura}, as a means of obviating some of the most recalcitrant problems that stand in the way of the implementation of large scale quantum computing: algorithm design \cite{behrman2008}, noise and decoherence \cite{behrman1, behrman2}, and scaleup \cite{behrmansteck}. Amalgamating artificial neural networks with intrinsic properties of quantum computing enables the QANN models to evolve as promising alternatives to quantum algorithmic computing~\cite{nam, puro}. Recent advances in both hardware and theoretical development may enable their implementation on the noisy intermediate scale (NISQ) computers that will soon be available~\cite{tachino, behrman, behrman1, behrman2}.  Konar \emph{et al.}\cite{konar1, konar3}, Schutzhold \emph{et al.}~\cite{schutz}, Trugenberger \emph{et al.}~\cite{trugenber} and Masuyama \emph{et al.}~\cite{loo} suggested quantum neural networks for pattern recognition tasks which deserve special mention for their contribution on QNN. \\
The classical self-supervised neural network architectures employed for binary image segmentation suffer from slow convergence problems~\cite{ghosh, bhatt2}. To overcome these challenges, the authors proposed the quantum version of the classical self-supervised neural network architecture relying on \emph{qubits} for faster convergence and accurate image segmentation and implemented on classical systems~\cite{konar1, bhatt1, konar3}. Furthermore, the recently modified versions of the network architectures relying on \emph{qubits} and characterized by multi-level activation function~\cite{konar4,konar5,konar6}, are also validated on MR images for brain lesion segmentation and reported promising outcome while compared with current deep learning architectures. However, the implementation of these quantum neural network models on classical systems is centred on the bi-level abstraction of \emph{qubits}. In most physical implementations, the quantum states are not inherently binary~\cite{gokhale}; thus, the \emph{qubit} model is only an approximation that suppresses higher-level states. The \emph{qubit} model can lead to slow and untimely convergence and distorted outcomes. Here, three-level quantum states or \emph{qutrits} (generally $D$-level \emph{qudits}) are introduced to improve the convergence of the self-supervised quantum network models.
\subsection{Motivation}
\label{inspiration}
The motivation behind the proposed QFS-Net over the deep learning based brain tumor segmentation~\cite{zikic1, olaf, pereira, wang} are as follows: 
\begin{enumerate}
    \item Huge volumes of annotated medical images are required for suitable training of a convolutional neural network, and it is also a paramount task to acquire. 
    \item The extensive and time-consuming training of deep neural network-based MR image segmentation requires high computational capabilities (GPU) and memory resources.   
    \item In contrast to automatic brain lesion segmentation, the slow convergence and over-fitting problems often affect the outcome, and hence extra efforts are required for suitable tuning of hyper-parameters of the underlying deep neural network architecture. 
    \item Moreover, the lack of image-specific adaptability of the convolutional neural network leads to a fall in accuracy for unseen medical image classes. 
\end{enumerate}
 A potential solution to devoid the requirement of training data and the problems faced by intensely supervised convolutional neural networks prevalent to medical image segmentation is a fully self-supervised neural network architecture with minimum human intervention. The novel qutrit-inspired fully self-supervised quantum learning model incorporated in the QFS-Net architecture presented in this article is a formidable contribution in exploiting the information of the brain lesions and poses a new horizon of research and challenges. 
 
\section{Fundamentals of Quantum Computing}
\label{QC}
Quantum computing offers the inherent features of superposition, coherence, decoherence, and entanglement of quantum mechanics in computational devices and enables implementation of quantum computing algorithms~\cite{ventura}. Physical hardware in classical systems uses binary logic; however, most quantum systems have multiple ($D$) possible levels. States of these systems are referred to as \emph{qudits}.
\subsection{Concept of Qudits}
\label{QC:qubits}In contrast to a two-state quantum system, described by a \emph{qubit}, a ($D>2$) multilevel quantum system is represented by $D$ basis states. We choose, as is usual the so-called ``computational'' basis:  $|0\rangle, |1\rangle, |2\rangle, \ldots |D-1\rangle$. A general pure state of the system is a superposition of these basis states represented as
\begin{equation}
|\psi\rangle=\alpha_0 |0\rangle + \alpha_1  |1\rangle + \alpha_2|2 \rangle + \ldots + \alpha_{D-1}|D-1 \rangle = \left[ {{\begin{array}{*{20}c}
		{\alpha_0 } \hfill \\
		{\alpha_1} \hfill \\
		\ldots \\
		{\alpha_{D-1}} \hfill \\
		\end{array} }} \right]
\end{equation}
subject to the normalization criterion $|\alpha_0|^2 + |\alpha_1|^2 + \ldots + |\alpha_{D-1}|^2 = 1$ where, $\alpha_0, \alpha_1, \ldots \alpha_{D-1}$ are complex quantities, i.e., $\{\alpha_i\} \in \mathbb{C}.$ Physically, the absolute magnitude squared of each coefficient $\alpha_i$ represents the probability of the system being measured to be in the corresponding basis state $|i\rangle$. \\
In this article, we use  a three-level system ($D=3$), i.e., a basis of $\{|0\rangle$, $|1\rangle$ and $|2\rangle\}$ for each quantum trit or \emph{qutrit}. One physical example of a \emph{qutrit} is a spin-1 particle. A general pure (coherent) state of a \emph{qutrit}~\cite{gokhale} is a superposition of all the three basis states, which can be represented as
\begin{equation}
|\psi\rangle=\alpha_0 |0\rangle + \alpha_1 |1\rangle + \alpha_2|2 \rangle = \left[ {{\begin{array}{*{20}c}
		{\alpha_0 } \hfill \\
		{\alpha_1 } \hfill \\
		{\alpha_2} \hfill \\
		\end{array} }} \right]
\end{equation}
subject to the normalization criterion $|\alpha_0|^2 + |\alpha_1|^2 + |\alpha_2|^2 = 1$. For example, the state
\begin{equation}
|\psi_3\rangle= \frac{2}{\sqrt{10}}|0\rangle + \frac{\sqrt{3}}{\sqrt{10}}|1\rangle + \frac{\sqrt{3}}{\sqrt{10}}|2\rangle
\end{equation}
has a probability of the being measured to be in the basis state $|2\rangle$ of
\begin{equation}
|\langle 2|\psi_3\rangle|^2 = \frac{3}{10}
\end{equation}
Similarly, the probabilities of the quantum state $|\psi_3\rangle$ being measured to be in each of the other two basis states $|0\rangle$ and $|1\rangle$  are $\frac{4}{10}$ and $\frac{3}{10}$ respectively.

\subsection{Quantum Operators}
\label{QC:CNOT}
We define generalized Pauli operators on \emph{qudits} as
\begin{equation}
X= \sum_{k=0}^{D-1} |k+1(mod 3)\rangle \langle k|,~~Z=\sum_{k=0}^{D-1}\theta^k |k\rangle \langle k|
\end{equation}
where $\theta = e^{\frac{2\pi}{D}}$ is the $D^{th}$ complex root of unity. That is, the operator $X$ shifts a computational basis state $|k\rangle$ to the next state, and the $Z$ operator multiplies a computational basis state by the appropriate phase factor. Note that these two operators generate the generalized Pauli group.

The Hadamard gate is one of the basic constituents of quantum algorithms, as its action creates superposition of the basis states. On \emph{qutrits} it is defined as~\cite{corbaci}
\begin{equation}
\mathcal{H} = \frac{1}{\sqrt{3}}\left[ {{\begin{array}{*{20}c}
		1 & 1 & 1  \hfill \\
		1 & {e^{\jmath \frac{2\pi}{3}}} & {e^{-\jmath \frac{2\pi}{3}}}  \hfill \\
		1 & {e^{-\jmath \frac{2\pi}{3}}} & {e^{\jmath \frac{2\pi}{3}}} \hfill \\
		\end{array} }} \right]
\end{equation}
The special case of the generalized Hadamard gate for \emph{qudits} is given by
\begin{equation}
\mathcal{H}|k\rangle = \sum_{i=0}^{D-1}\theta_k^i|i\rangle, where~~
\theta_i^D = \cos(\frac{2i\pi}{D}) + \jmath \sin(\frac{2i\pi}{D}) = e^{\frac{2i\pi}{D}}
\end{equation}
Here $\jmath$ is the imaginary unit and the angle $\theta$ is the $i^{th}$ root of 1. \\
We define a rotation gate  $\mathcal{R}(\omega) = e^{\frac{\jmath\omega}{3}}$, which transforms a \emph{qutrit} in state ($\alpha_0, \alpha_1, \alpha_2 $) to the (rotated) state ($\alpha_0', \alpha_1', \alpha_2'$), as follows:
\begin{equation}
{\begin{array}{*{20}c}
	\left[ {{\begin{array}{*{20}c}
			{\alpha_0'} \hfill \\
			{\alpha_1'} \hfill \\
		    {\alpha_2'} \hfill 
			\end{array} }} \right] \hfill 	=\frac{1}{2} \left[ {{\begin{array}{*{20}c}
		1+\cos\omega & -\sqrt{2}\sin\omega & 1-\cos\omega  \hfill \\
		\sqrt{\sin\omega} & 2\cos\omega & -\sqrt{2}\sin\omega  \hfill \\
		1-\cos\omega & \sqrt{2}\sin\omega & 1+\cos\omega \hfill \\
		\end{array} }} \right]\times \left[ {{\begin{array}{*{20}c}
			{\alpha_0} \hfill \\
			{\alpha_1} \hfill \\
		    {\alpha_2} \hfill
			\end{array} }} \right] \hfill
	\end{array} }
\end{equation}
Note that the rotation gate defined above is a unitary operator. 

\section{Quantum Neural Network Model based on Qudits (QNNM)}
\label{QINN:Add}
A quantum neural network dealing with discrete data is realized on a classical system using quantum algorithms and acts on quantum states through a layered architecture. In this proposed \emph{qudit} embedded quantum neural network model, the classical network inputs are converted into $D$-dimensional quantum states $[0, \frac{2\pi}{D}]$ or \emph{qudits}. Let the $k^{th}$  input be given by  $x_k$. We apply a standard classical sigmoid activation function $f_{QNNM} (x_k)$, which yields binary classical outcome $[0, 1]$.
\begin{equation}
\begin{split}
f_{QNNM} (x_k) = \frac{1}{1+ e^{-x_k}} 
\end{split}
\end{equation} 
We then map that quantity onto the amplitude for the $k^{th}$ basis state as
\begin{equation}
\label{Eq:sig}
\begin{split}
|\alpha_k\rangle =  (\frac{2\pi}{D}f_{QNNM}(x_k)) 
\end{split}
\end{equation}

The suggested QNNM model comprises multiple $D$-dimensional \emph{qudits}, $\mathbb{Z}_D = \{(|z_1\rangle, |z_2\rangle, |z_3\rangle,\ldots |z_D\rangle)^T : |z_k\rangle \in \mathbb{Z}(k=1,2,\ldots D)\}$. The inner product between the input quantum states $|\psi_D\rangle= (|\alpha_1\rangle, |\alpha_2\rangle, |\alpha_3\rangle, \ldots ,|\alpha_D\rangle)^T$ and the quantum weights $|W_D\rangle= (|\theta_1\rangle, |\theta_2\rangle, |\theta_3\rangle, \ldots, |\theta_D\rangle)^T$ is defined as 
\begin{equation}
\begin{split}
    \langle \psi_D|W_D\rangle = \sum_{k=1}^{D}{\langle\alpha_k}{|\theta_k\rangle} =  \mathcal{T}_D|\alpha_k\rangle \mathcal{H}_D|\theta_k\rangle = \\ \sum_{k=1}^{D}\alpha_k|k\rangle (\sum_{k=1}^{D} \cos(\frac{2k\pi}{D}) + \jmath \sin(\frac{2k\pi}{D}))
\end{split}
\end{equation}
where $\mathcal{T}_D$ and $\mathcal{H}_D$ are the transformation (realization mapping) and the Hadamard gate, respectively. $\bar{\psi_D}$ is the complex conjugate of $\psi_D$ and is defined as
\begin{equation}
\begin{split}
    \langle \psi_D| = |\psi_D\rangle^\dagger = (|\bar{\alpha_1}\rangle,|\bar{\alpha_2}\rangle,|\bar{\alpha_3}\rangle, \ldots ,|\bar{\alpha_D}\rangle) 
\end{split}
\end{equation}
In the suggested quantum neural network model, let us consider the set of all quantum states be denoted as $Q_D(\mathbb{Z})$ and the $D$-dimensional realization transformation, $\mathcal{T}_D: Q_D(\mathbb{Z}) \rightarrow \mathbb{R}^{2D}$ is defined as
\begin{equation}
\begin{split}
\mathcal{T}|\psi_D\rangle = \\(Re|\alpha_1\rangle, Im|\alpha_1\rangle,Re|\alpha_2\rangle, Im|\alpha_2\rangle, \ldots Re|\alpha_D\rangle, Im|\alpha_D\rangle)^T 
\end{split}
\end{equation}
for all $|\psi_D\rangle = (|\alpha_1\rangle, |\alpha_2\rangle, \ldots |\alpha_D\rangle)^T \in Q_D(\mathbb{Z})$ and $\forall i\in D, |\alpha_i\rangle = \cos{\omega_i}|0\rangle + j \sin{\omega_i}|1\rangle$. The input-output association of a $D$-dimensional basic quantum neuron in the proposed QNNM model at a particular epoch ($t$) is modeled as
\begin{equation}
\begin{split}
|\mathcal{O}^t_k\rangle= \mathcal{T_D}(|h^t_k\rangle) = \mathcal{T}(\frac{2\pi}{D} \delta_{Dk}^t - \arg(|y_k^t\rangle))
\end{split}
\end{equation}
where,
\begin{equation}
\label{eq:inp-out}
\begin{split}
|y^t_k\rangle= \sum_{i=1}^N \mathcal{H_D}(|\theta^t_{i,k}\rangle) \mathcal{T_D}(|\mathcal{O}^{t-1}_k\rangle) - \mathcal{H_D}(|\xi^t_i\rangle)
\end{split}
\end{equation}
Here, the quantum phase transformation parameter (weight) between the $k^{th}$ output neuron and the $i^{th}$ input neuron is $|\theta_{i,k}\rangle$ and the activation is $|\xi^t_i\rangle$. The $D$-dimensional Hadamard gate parameters are designated by $\delta_{Dk}$. Considering the basis state $|D-1\rangle$, the true outcome of the quantum neuron $k$ at the output layer is obtained through quantum measurement of $D$-dimensional quantum state $|\mathcal{O}^t_k\rangle$  as
\begin{equation}
\mathcal{M}^k_{QNMM} = |Im(|\mathcal{O}_k^{t}\rangle)|^2
\end{equation}
where, the imaginary section of $\mathcal{O}^t_k$ is referred to as $Im(\mathcal{O}_k^{t})$.\\
It is worth noting that the realization mapping $\mathcal{T}$ transforms quantum states to probability amplitudes and hence the quantum state is destroyed on implementation in classical systems. However, the suggested quantum neural network model is not a quantum neural network in the true sense of the term. It is a quantum mechanics-inspired hybrid neural network model implementable on  classical systems. 

\section{Quantum Fully Self-supervised Neural Network (QFS-Net)}
\label{QFS-Net}

\begin{figure*}
	\centering
	\includegraphics[scale=0.4]{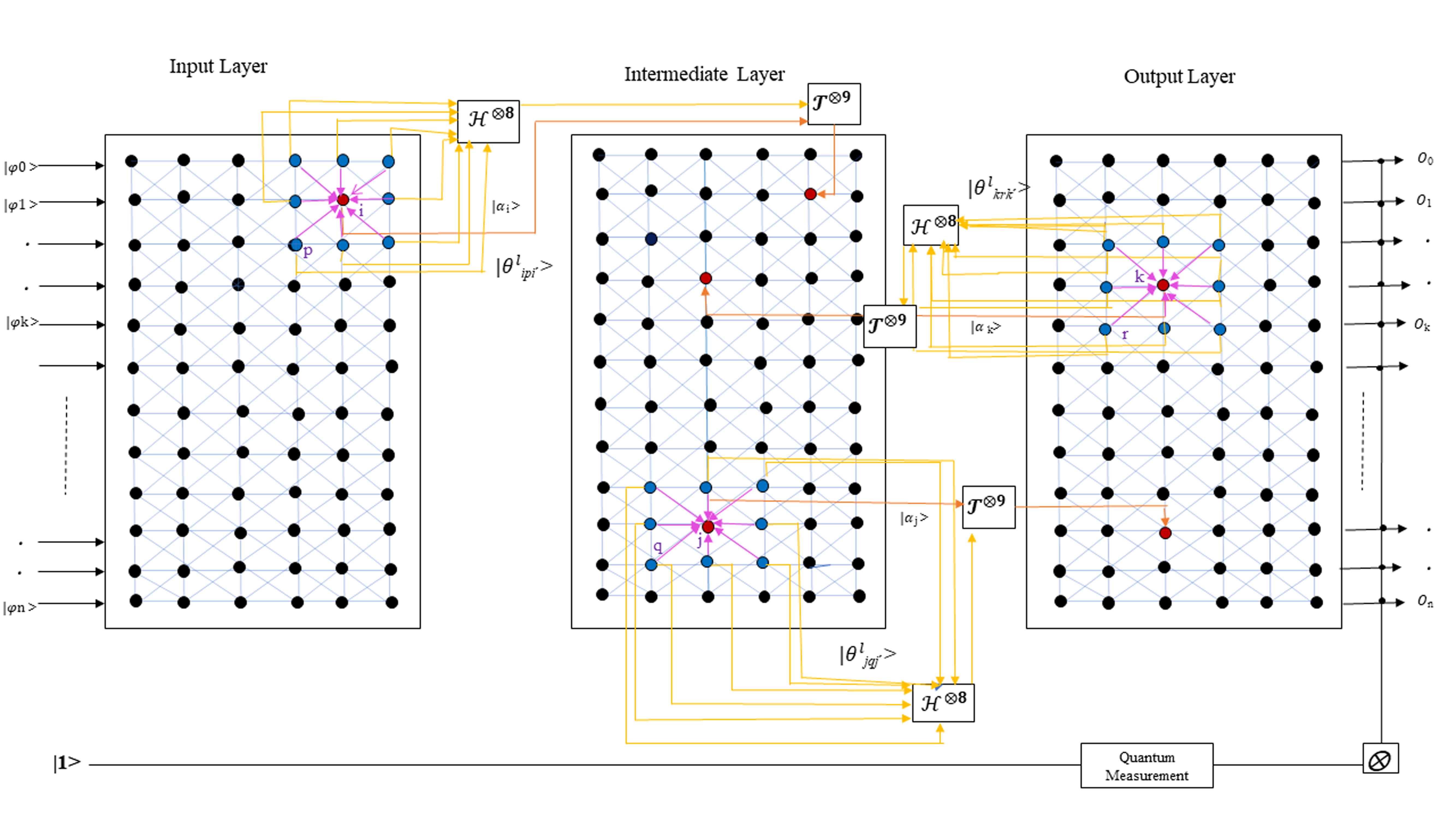}
	\caption{qutrit-inspired Quantum Fully Self-Supervised Neural Network (QFS-Net) architecture where $\mathcal{H}$ represents Hadamard gate and $\mathcal{T}$ is realization gate (only three inter-layer connections are shown for clarity).}
	\label{fig:QBDSONN}
\end{figure*}

The suggested quantum fully self-supervised neural network architecture comprises trinity layers of \emph{qutrit} neurons arranged as input, intermediate and output layers. A schematic outline of the QFS-Net architecture as a quantum neural network model is illustrated  Figure~\ref{fig:QBDSONN}.
The information processing unit of the QFS-Net architecture is depicted using quantum neurons (\emph{qutrits}) reflected in the trinity layers using the combined matrix notation. 
\[\left[
\begin{array}{ccccc}
|\psi_{11}\rangle & |\psi_{12}\rangle & |\psi_{13}\rangle & \ldots & |\psi_{1m}\rangle \\
\ldots & \ldots & \ldots & \ldots & \ldots \\
\ldots & \ldots & \ldots & \ldots & \ldots \\
\ldots & \ldots & \ldots & \ldots & \ldots \\
|\psi_{n1}\rangle & |\psi_{n2}\rangle & |\psi_{n3}\rangle & \ldots & |\psi_{nm}\rangle \\
\end{array}
\right]\]
Hence, each quantum neuron constitutes a \emph{qutrit} state designated as $\psi_{ij}$.

Each layer of the quantum self-supervised neural network architecture is organized by combining the \emph{qutrit} neurons in a fully-connected fashion with intra-connection strength as $\frac{2\pi}{3}$ (\emph{qutrit} state). The main characteristic of the network architecture lies in the organization of the 8-connected second-order neighborhood subsets of each quantum neuron in the layers of the underlying architecture and propagation to the subsequent layers for further processing. The input, intermediate/hidden and output layers are inter-connected through self-forward propagation of the \emph{qutrit} states in the 8-connected neighborhood fashion. On the contrary, the inter-connections are established from the output layer to intermediate layer entailing self-counter-propagation obviating the quantum back-propagation algorithm and thereby reducing time complexity. 
Finally, a quantum observation process allows the \emph{qutrit} states to collapse to one of the basis states ($0$ or $1$ as $2$ is considered as a temporary state). We obtain true outcome at the output layer of the QFS-Net once the network converges, else quantum states undergo further processing.  

\subsection{Qutrit-inspired Fully Self-supervised Quantum Neural Network Model}
\label{QISNNM}
The novel quantum fully self-supervised neural network model based on \emph{qutrits} adopts twofold schemes. The \emph{qutrit} neurons of each layer are realized using a $\mathcal{T}$ transformation gate (realization mapping) and the inter-connection weights are mapped using the phase Hadamard gates ($\mathcal{H}$) applicable on \emph{qutrits}. The angle of rotation is set as relative difference of quantum information (marked by pink arrow in Figure~\ref{fig:QBDSONN}) between each candidate \emph{qutrit} neuron and the neighborhood \emph{qutrit} neuron of the same layer employed in the rotation gate for updating the inter-layer interconnections. The rotation angle for the inter-connection weights and the threshold are set as $\omega$ and $\gamma$, respectively. The inter-connection weights between the \emph{qutrit} neurons (denoted as $k$ and $i$) of two adjacent layers are depicted as $|\theta_{ik}\rangle$ and measured as the relative difference between the $i^{th}$ candidate \emph{qutrit} neuron and the 8-connected neighborhood quantum neuron $k$. The realization of the network weights are mapped using the Hadamard gate ($\mathcal{H}$) inspired by the proposed QNNM model by suppressing the highest basic level ($|2\rangle$) of \emph{qutrit} as a temporary storage as
\begin{equation}
\mathcal{H}(|\theta_{ik}\rangle) = \cos (\frac{2\pi}{3}\omega_{i,k}) + \jmath \sin (\frac{2\pi}{3}\omega_{i,k}) =\left[
\begin{array}{c}
\cos (\frac{2\pi}{3}\omega_{i,k}) \\
\sin (\frac{2\pi}{3}\omega_{i,k}) \\
\end{array}
\right]
\end{equation}
where, $\jmath$ is an imaginary unit. The role of relative measure of the quantum fuzzy information lies in the fact that the distinction between the foreground and background image pixels is clearly visible on adapting the relative measures. Assuming the quantum fuzzy grade information at the $i^{th}$ candidate neuron and its 8-connected second order neighborhood neuron as $\mu_i$ and $\mu_{i,k}$ respectively, the angle of the Hadamard gate is determined as
\begin{equation}
\label{eq:alpha}
\begin{split}
\omega_{i,k} = 1-(\mu_i-\mu_{i,k}); k \in \{1, 2, \ldots 8\}
\end{split}
\end{equation}
The 8-fully intra-connected spatially arranged neighborhood \emph{qutrit} neurons contribute to the candidate quantum neuron (say $i'$) of the adjacent layer through the transformation gate ($\mathcal{T}$) and the realization mapping defined as 
\begin{equation}
\begin{split}
|\psi_{i'}\rangle = \sum_{k}\mathcal{T}(|\mu_{i,k}\rangle)\mathcal{H}(|\theta_{iki'}\rangle) = \\ \sum_{k}[\mu_{i,k}\{\cos (\frac{2\pi}{3}\omega_{i,k}) + \jmath \sin (\frac{2\pi}{3}\omega_{i,k})\}]
\end{split}
\end{equation}
In addition, the contribution of the 8-fully intra-connected spatially arranged neighborhood \emph{qutrit} neurons are accumulated at the candidate \emph{qutrit} neuron as the quantum fuzzy context sensitive activation ($\xi_i$) and is presented using the Hadamard gate as
\begin{equation}
\mathcal{H}(|\xi_i\rangle) =\cos (\frac{2\pi}{3}\gamma_i) + \jmath \sin (\frac{2\pi}{3}\gamma_i)=\left[
\begin{array}{c}
\cos (\frac{2\pi}{3}\gamma_i) \\
\sin (\frac{2\pi}{3}\gamma_i) \\
\end{array}
\right]
\end{equation}
where, the angle of the Hadamard gate is defined as
\begin{equation}
\label{eq:gamma}
\begin{split}
\gamma_i = (\sum_{k}\mu_{i,k})
\end{split}
\end{equation}
The self-supervised forward and counter propagation of the QFS-Net are guided by a novel \emph{qutrit} based adaptive multi-class Quantum Sigmoid (\emph{QSig}) activation function with quantum fuzzy context sensitive thresholding as discussed in the following subsection~\ref{QMUSIG}. The basis of network dynamics of the QFS-Net is centred on the bi-directional self-organized propagation of the \emph{qutrit} states between the intermediate and output layers via updating of inter-connection links.\\
The network basic input-output relation is presented through the composition of a sequence using the transformation gate ($\mathcal{T}$) and the realization mapping defined as  
\begin{equation}
\begin{split}
|\psi^l_k\rangle =  QSig(\sum_{i=1}^8 \mathcal{T}(\psi^{l-1}_{k,i}) \mathcal{H}(\langle \theta^l_i|\xi^l_i \rangle))
\end{split}
\end{equation}
where, $|\psi^l_k\rangle$ is the output of the $k^{th}$ constituent \emph{qutrit} neuron at the $l^{th}$ layer and the contribution of each 8-connected neighborhood \emph{qutrit} neurons of the $k^{th}$ candidate neuron is  expressed as $|\psi^{l-1}_k\rangle$  i.e. ,
\begin{equation}
\begin{split}
|\psi^l_k\rangle = \mathcal{T}\left[\frac{2\pi}{3}\delta^l_k- \arg \{\sum_{i=1}^8\mathcal{H}(|\theta^l_{k,i}\rangle)\mathcal{T}(|\psi^{l-1}_{k,i}\rangle)-\mathcal{H}(|\xi^l_k\rangle)\}\right]
\end{split}
\end{equation}
Quantum observation on a \emph{qutrit} neuron transforms a quantum state into a basis state and a true outcome ($|1\rangle$) is obtained on measurement from the \emph{qutrit} neuron considering the imaginary section of $|\psi^l_k)$ as 
\begin{equation}
\begin{split}
\mathcal{O}^l_k= |Im(|\psi^l_k\rangle)|^2
\label{eq:qunatumobser}
\end{split}
\end{equation}
i.e
\begin{equation}
\begin{split}
\mathcal{O}^l_k =  QSig(\sum_{i=1}^8 \mathcal{T}(\psi^{l-1}_{k,i})\cos(\frac{2\pi}{3}(\omega^l_{k,i}-\gamma^l_k))+ \\ \jmath \sin(\frac{2\pi}{3}(\omega^l_{k,i}-\gamma^l_k)))
\end{split}
\end{equation}
where, the quantum phase transmission parameter from the input \emph{qutrit} neuron $i$ (the neighborhood of $k^{th}$ \emph{qutrit} neuron at the layer $l-1$ is depicted as $i$) to intermediate \emph{qutrit} neuron $k$ with activation $\xi^l_k$, is $\omega^l_{k,i}$. The rotation gate parameters are expressed as $\delta^l_k$ with the parameters of activation as $\gamma_k^l$ at the layer $l$. The activation function employed in the proposed QFS-Net model is a novel adaptive multi-class \emph{qutrit} embedded sigmoidal ($QSig$) activation function which is illustrated in the following subsection~\ref{QMUSIG}. 

\subsection{Qutrit-Inspired Self-supervised Learning of QFS-Net} 
\label{QFS-Net: learning}
Let us consider, the interconnection weights in terms of \emph{qutrit} between the input and the hidden or intermediate layer are expressed as $|\theta^l_{ipi'}\rangle$ (here any candidate \emph{qutrit} neuron at the input layer is $i$, its corresponding candidate neuron at the next subsequent intermediate layer is $i'$ and its corresponding 8-connected neighborhood neurons are described by $p$) and for the intermediate layer to the output layer are $|\theta^l_{jqj'}\rangle$ (here any candidate \emph{qutrit} neuron at the intermediate layer is $j$, its corresponding candidate neuron at the next subsequent output layer is $j'$ and its corresponding 8-connected neighborhood neurons are described by $q$) at the $l^{th}$ iteration. The activation at the intermediate layer and output layer are expressed as $|\xi^l_{j}\rangle$ and $|\xi^l_{k}\rangle$, respectively. The self-supervised counter-propagation of the quantum states from output to intermediate layer is performed through the interconnection weight $|\theta^l_{krk'}\rangle$ (here any candidate \emph{qutrit} neuron at the output layer is $k$, its corresponding candidate neuron at the next subsequent intermediate layer is $k'$ and its corresponding 8-connected neighborhood neurons are described by $r$). The outcome of a \emph{qutrit} neuron ($|\psi^l_k\rangle$) at the output layer can be expressed as
\begin{equation}
\begin{split}
|\psi^l_k\rangle= QSig(\sum_{q=1}^8\mathcal{T}(|\psi^{l-1}_{jq}\rangle)\mathcal{H}(\langle \theta^l_{jqj'}|\xi^l_k \rangle) = QSig(\sum_{q=1}^8 \\\mathcal{T}(\frac{2\pi}{3}\times QSig(\sum_{p=1}^8 (\frac{2\pi}{3} x_{ip}) \mathcal{H}(\langle \theta^{l-1}_{ipi'}|\xi^{l-1}_j \rangle))) \mathcal{H}(\langle\theta^l_{jqj'}|\xi^l_k \rangle)
\end{split}
\end{equation}
i.e.,
\begin{equation}
\begin{split}
|\psi^l_k\rangle= QSig(\sum_{q=1}^8 \mathcal{T}(\frac{2\pi}{3}\times(QSig(\sum_{p=1}^8 (\frac{2\pi}{3} x_{ip})\\ \cos(\frac{2\pi}{3}(\omega^l_{ipi'}-\gamma^l_j)) \cos(\frac{2\pi}{3}((\omega^l_{jqj'}-\gamma^l_k)) +  \\ \jmath \sin(\frac{2\pi}{3}(\omega^l_{ipi'}-\gamma^l_j)) \sin (\frac{2\pi}{3}(\omega^l_{jqj'}-\gamma^l_k)))))
\end{split}
\end{equation}
where, $x_{ip}$ represents the classical input to the neighborhood neuron $p$ with respect to a candidate neuron $i$ at the input layer which is subsequently transformed  to a \emph{qutrit} state ($|\phi_{ip}\rangle= \frac{2\pi}{3}x_{ip}$) and $\jmath$ is an imaginary unit. An adaptive multi-class \emph{qutrit} embedded sigmoidal (\emph{QSig}) activation function employed in this self-supervised network model governs the activation at the intermediate and output layers and also the subsequent processing of the quantum states guided by various thresholding schemes. 

\subsection{Adaptive Multi-class Quantum Sigmoidal (QSig) activation function}
\label{QMUSIG}
 In this paper, we have introduced an adaptive multi-class sigmoidal activation function in quantum formalism suitable for pixel wise multi-class segmentation of medical images varying with multi-intensity gray-scales. The proposed \emph{QSig} activation function is the modification on the recently developed quantum multi-level sigmoid activation function employed in authors' previous work~\cite{konar4,konar5}. An optimized version of similar function is also introduced in ~\cite{konar6}. However, the requirement of finding optimal thresholding of the images in the activation function is computationally exhaustive and time dependent. The proposed \emph{QSig} relies on an adaptive step length incorporating the total number of segmentation levels with various schemes of activation. The \emph{QSig} activation function, employed in the QFS-Net model is defined as
\begin{equation}
\label{eq:1}
QSig(x)=\frac{1}{\kappa_\vartheta+e^{-\lambda(xh-\eta)}}
\end{equation}
where, $QSig (x)$ represents the adaptive multi-class quantum sigmoidal (\emph{QSig}) activation function with steepness parameter $\lambda$, step size $h$ and activation $\eta$ described by \emph{qutrits}. The multi-level class output, $\kappa_\vartheta$ as \emph{qutrit} is defined as
\begin{equation}
\kappa_\vartheta= \frac{Q_N}{\tau_\vartheta-\tau_{\vartheta-1}}
\end{equation}
The gray-scale intensity index is expressed as $\kappa_\vartheta$ ($1\leq \kappa_\vartheta \leq L$) where $\vartheta$ is the class index. The $\vartheta^{th}$ and ${\vartheta-1}^{th}$ class responses are denoted as  $\tau_\vartheta$ and $\tau_{\vartheta-1}$, respectively and the sum of the containment of $8$-connected neighborhood \emph{qutrit} neurons representing gray-scale pixels is denoted by $Q_N$. 
\begin{figure}[htbp]
\centering
\centering
\subcaptionbox{$L=3$}{\includegraphics[width=1.5in]{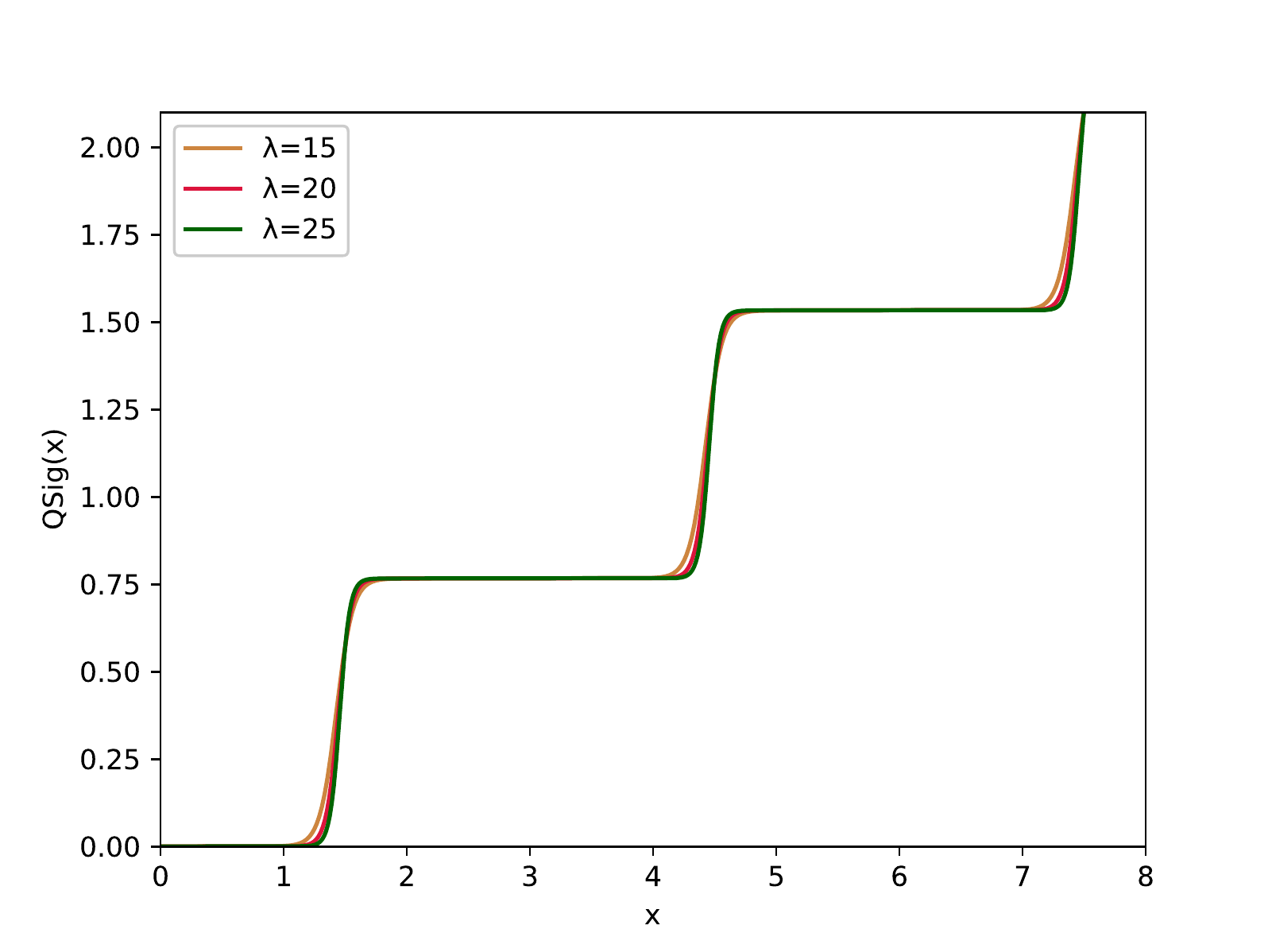}}
 \subcaptionbox{$L=4$}{\includegraphics[width=1.5in]{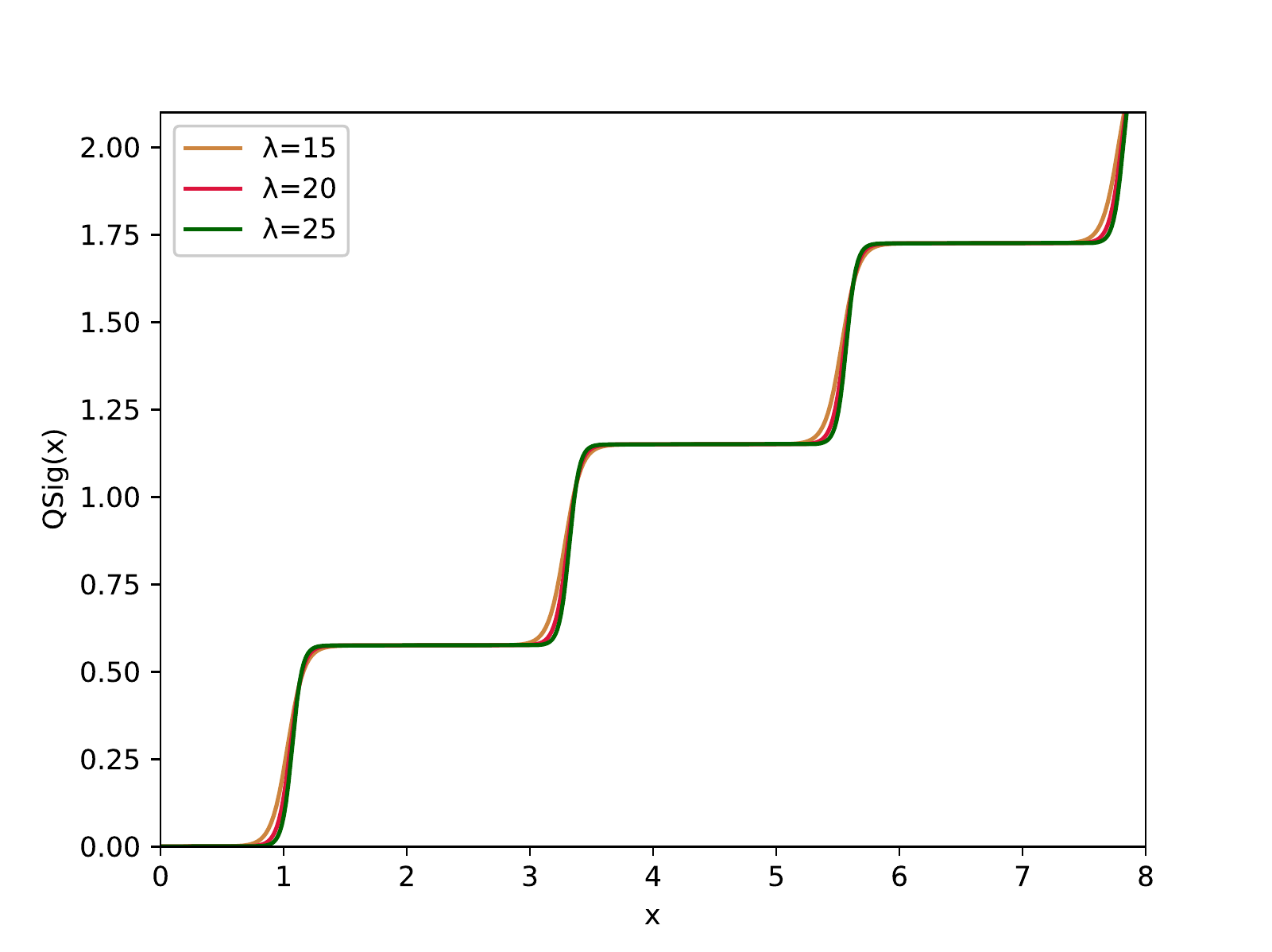}}
 \subcaptionbox{$L=6$}{\includegraphics[width=1.5in]{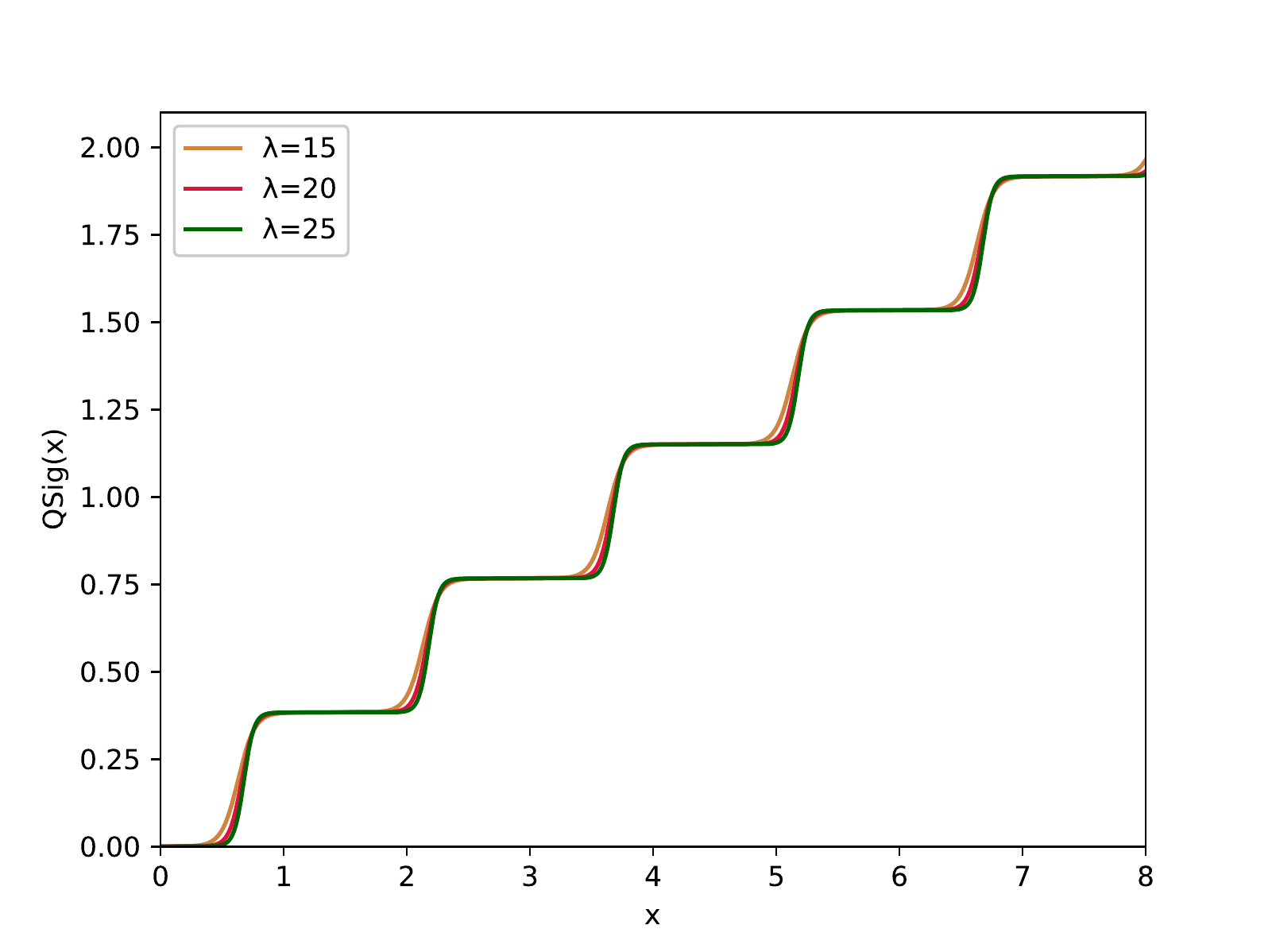}}
 \subcaptionbox{$L =8$}{\includegraphics[width=1.5in]{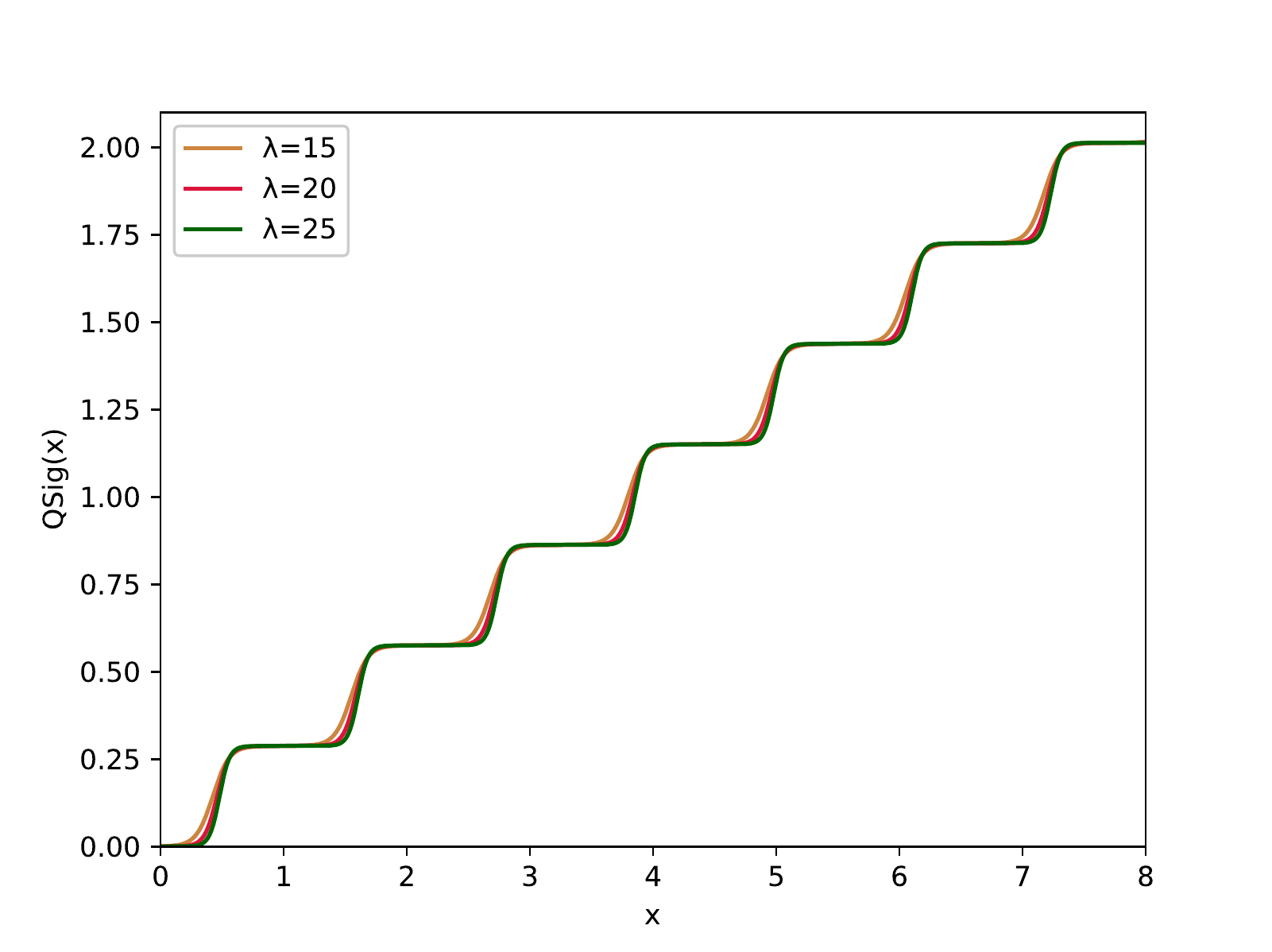}}
\caption{Multi-level class outcome of \emph{QSig} activation function for $\lambda=15, 20, 25$ and $h=1$ with segmentation levels}
	\label{fig:QMUSGPlot}
\end{figure}
The generalized version of the \emph{QSig} activation function defined in Equation~\ref{eq:1} can be modified leveraging $\kappa_\vartheta$ with various subnormal responses $\sigma_{\kappa_\vartheta}$ as \emph{qutrit} where  $ 0 \leq \sigma_{\kappa_\vartheta} \leq \frac{2\pi}{3}$. The multi-level class output is obtained on superposition of the subnormal responses and the generic \emph{QSig} activation function can be expressed as
\begin{equation}
\label{eq:2}
QSig(x;\kappa_\vartheta, \tau_\vartheta)=\frac{1}{\kappa_\vartheta+e^{-\lambda(x-(\vartheta-\frac{L+1}{2})\tau_{\vartheta-1}-\eta)}}
\end{equation}
In order to ensure that the number of distinct $\kappa_\vartheta$ parameters is to be equal to the number of multi-level classes ($L-1$), Equation~\ref{eq:3} depicts the closed form of the resultant \emph{QSig} function as 
\begin{equation}
\label{eq:3}
\begin{split}
Qsig_{R}(x)=\sum_{\vartheta=1}^{L} Qsig(x- (\vartheta-\frac{L+1}{2})\tau_{\vartheta-1});\\ (\vartheta-\frac{L+1}{2}) \tau_{\vartheta-1} \leq x \leq \vartheta \tau_\vartheta
\end{split}
\end{equation}
Substituting Equation~\ref{eq:2} in Equation~\ref{eq:3}, the updated form is expressed as
\begin{equation}
\label{eq:4}
QSig_{R}(x;\kappa_\vartheta, \tau_\omega)=\sum_{\vartheta=1}^{L} \frac{1}{\kappa_\vartheta+e^{-\lambda(x-(\vartheta-\frac{L+1}{2})\tau_{\vartheta-1}-\eta)}}
\end{equation}
Different forms of the \emph{QSig} activation function with different values of the steepness parameters are illustrated in Figure\ref{fig:QMUSGPlot}. 

\subsection{Updating Inter-connection Weight using Hadamard Gate}
\label{QBDSONN:weight}

\begin{figure}[htbp]
  \centering
 \subcaptionbox{QIS-Net, S1}{\includegraphics[width=1.5in]{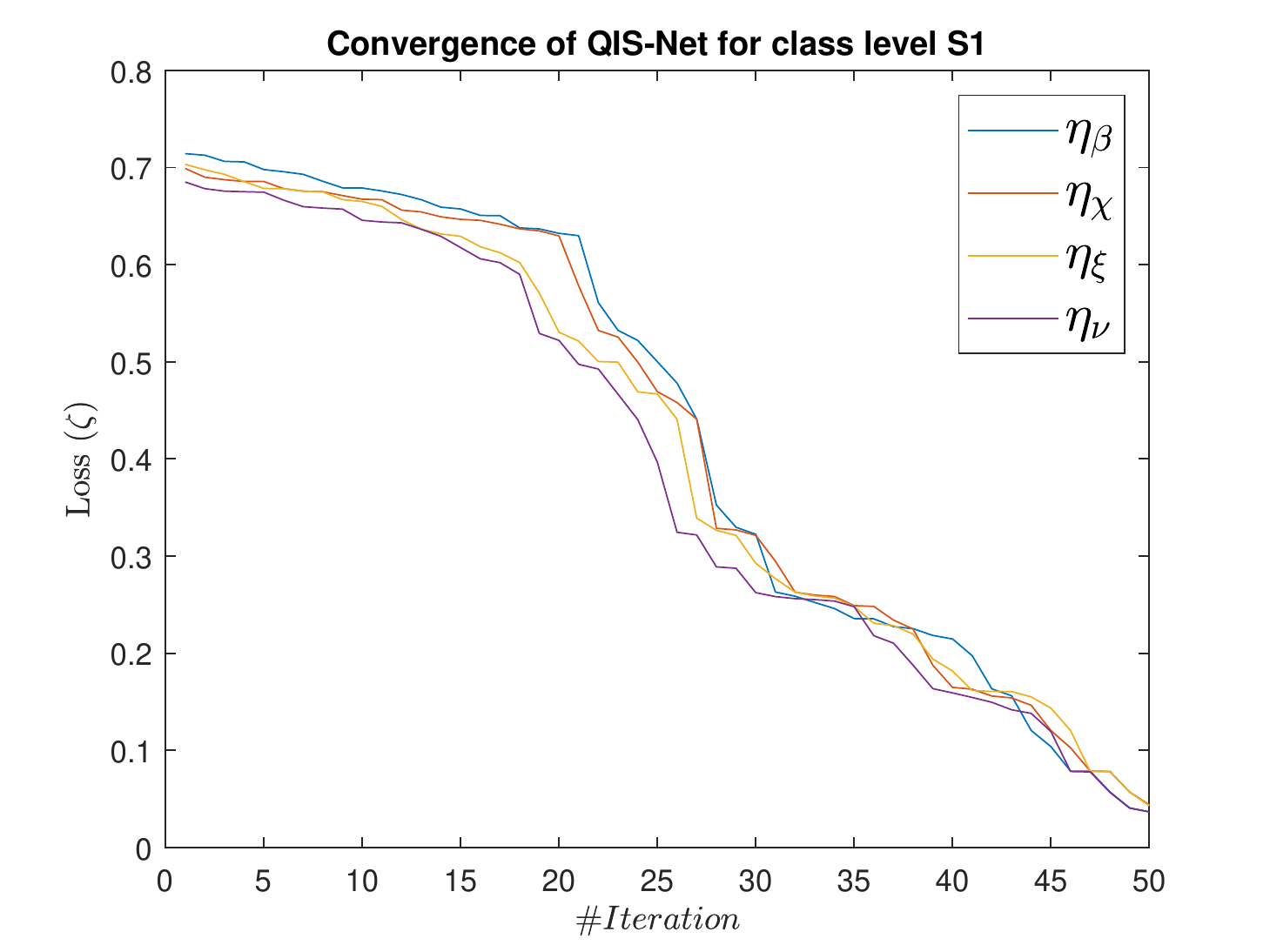}}
 \subcaptionbox{QIS-Net, S2}{\includegraphics[width=1.5in]{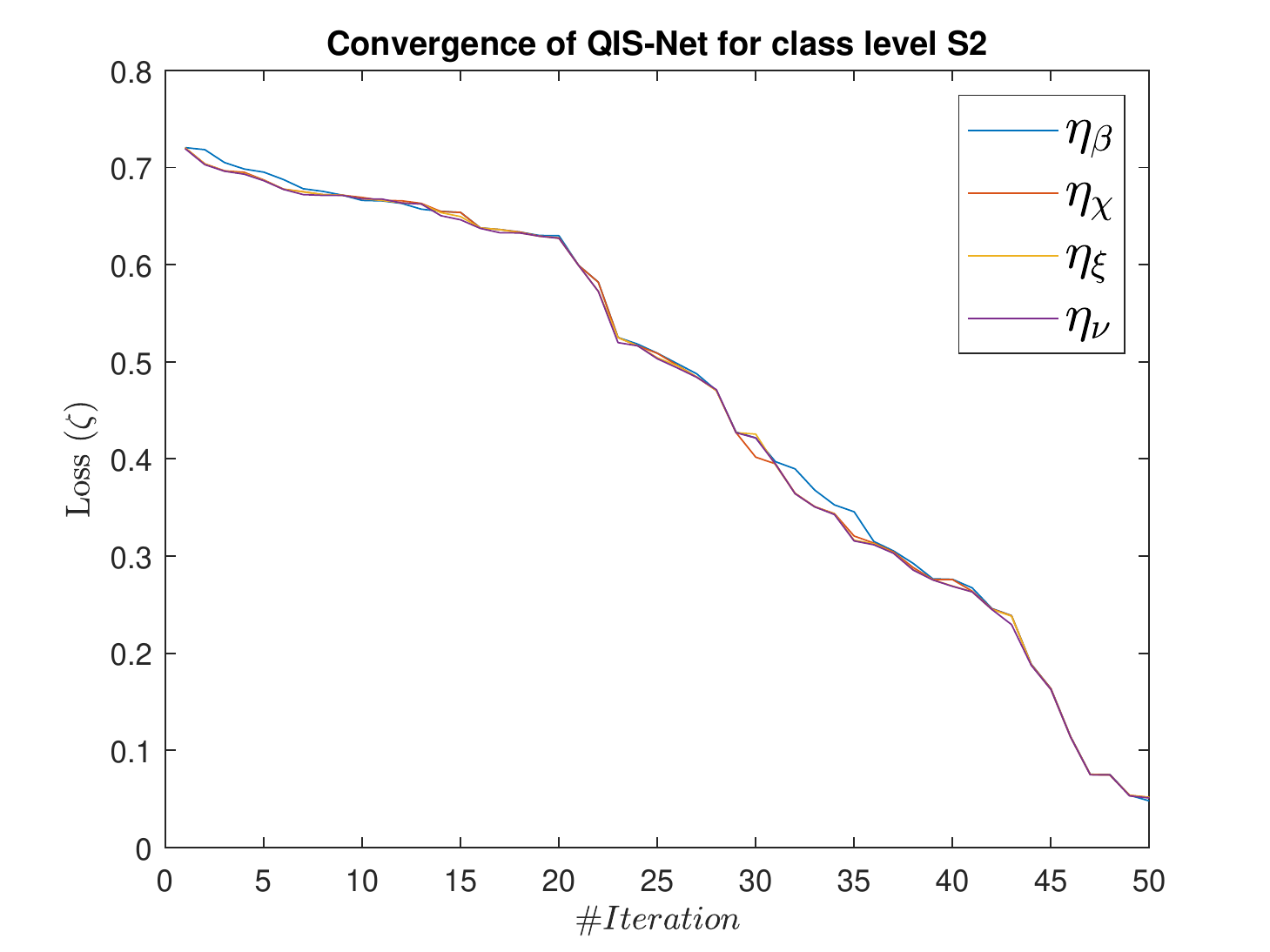}}
 \subcaptionbox{QIS-Net, S3}{\includegraphics[width=1.5in]{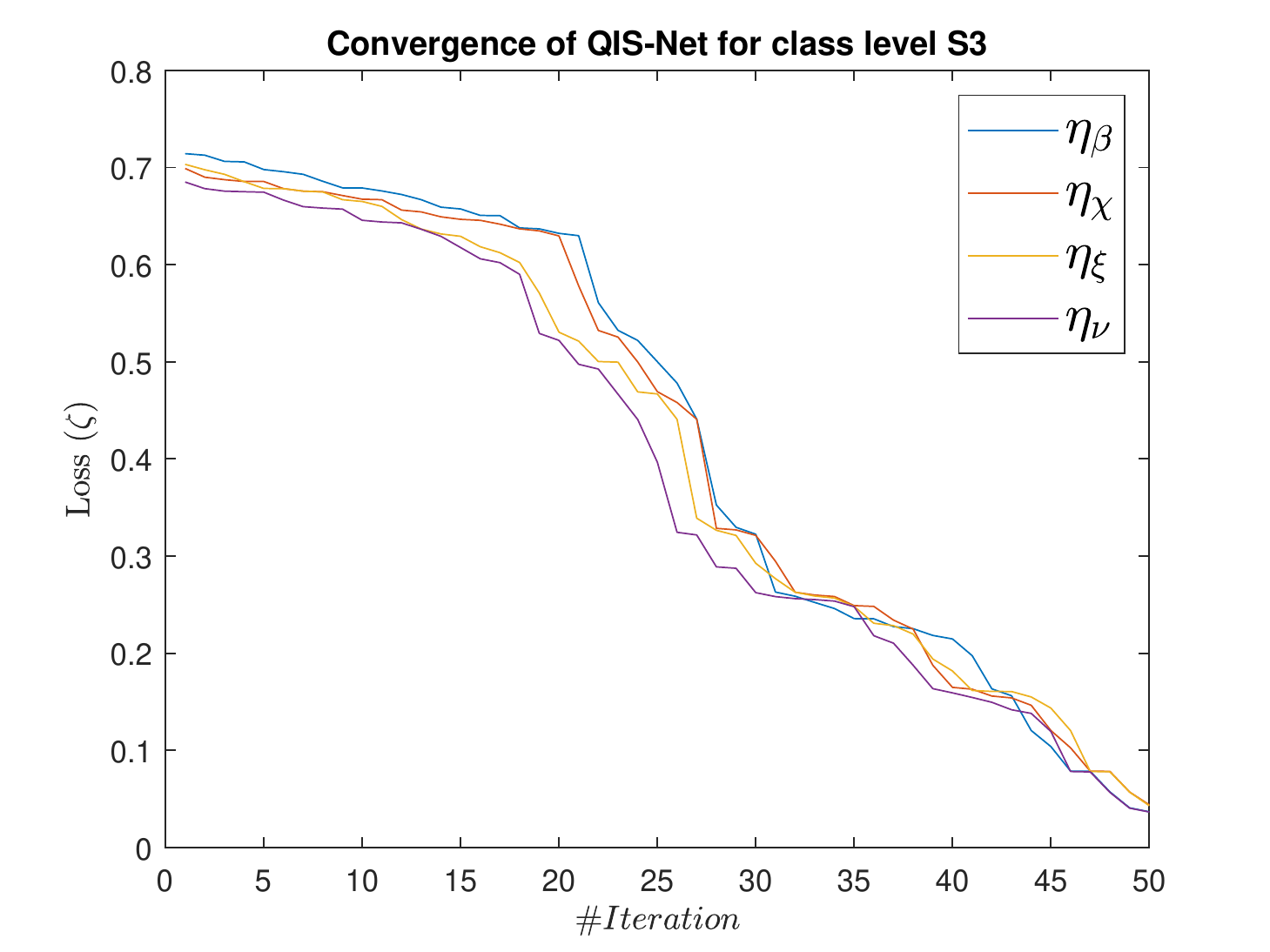}}
 \subcaptionbox{QIS-Net, S4}{\includegraphics[width=1.5in]{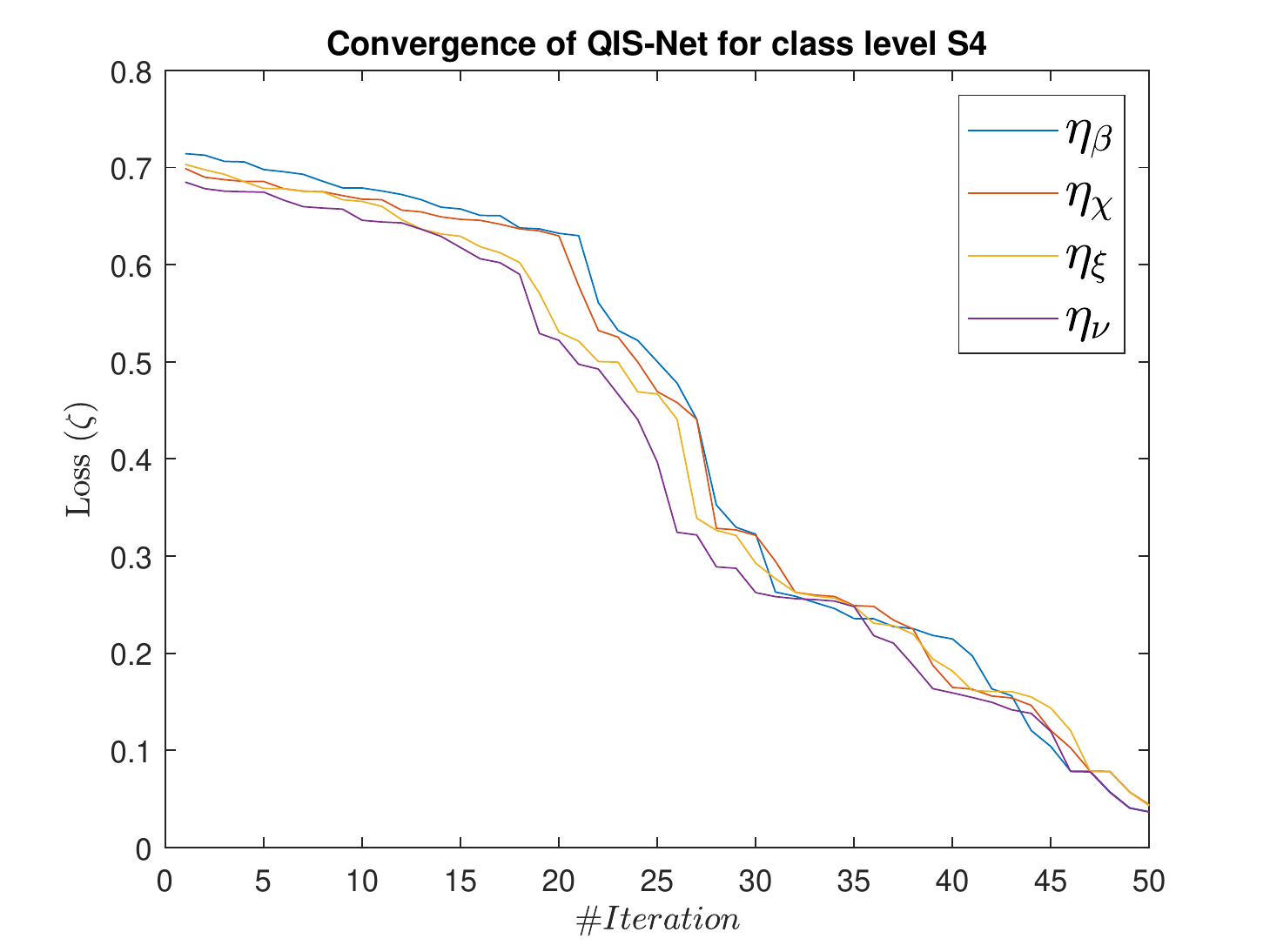}}
 \subcaptionbox{QFS-Net, S1}{\includegraphics[width=1.5in]{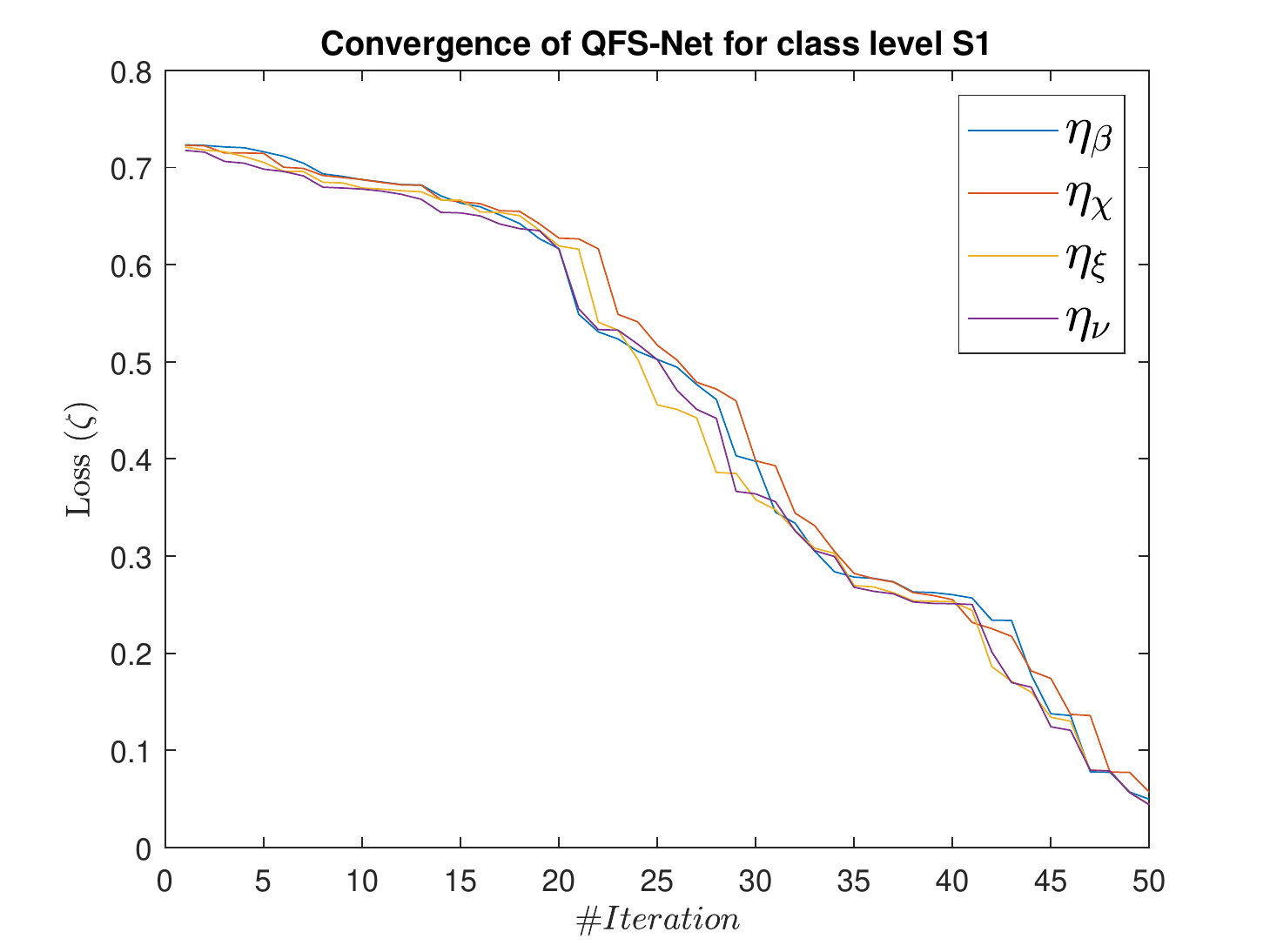}}
 \subcaptionbox{QFS-Net, S2}{\includegraphics[width=1.5in]{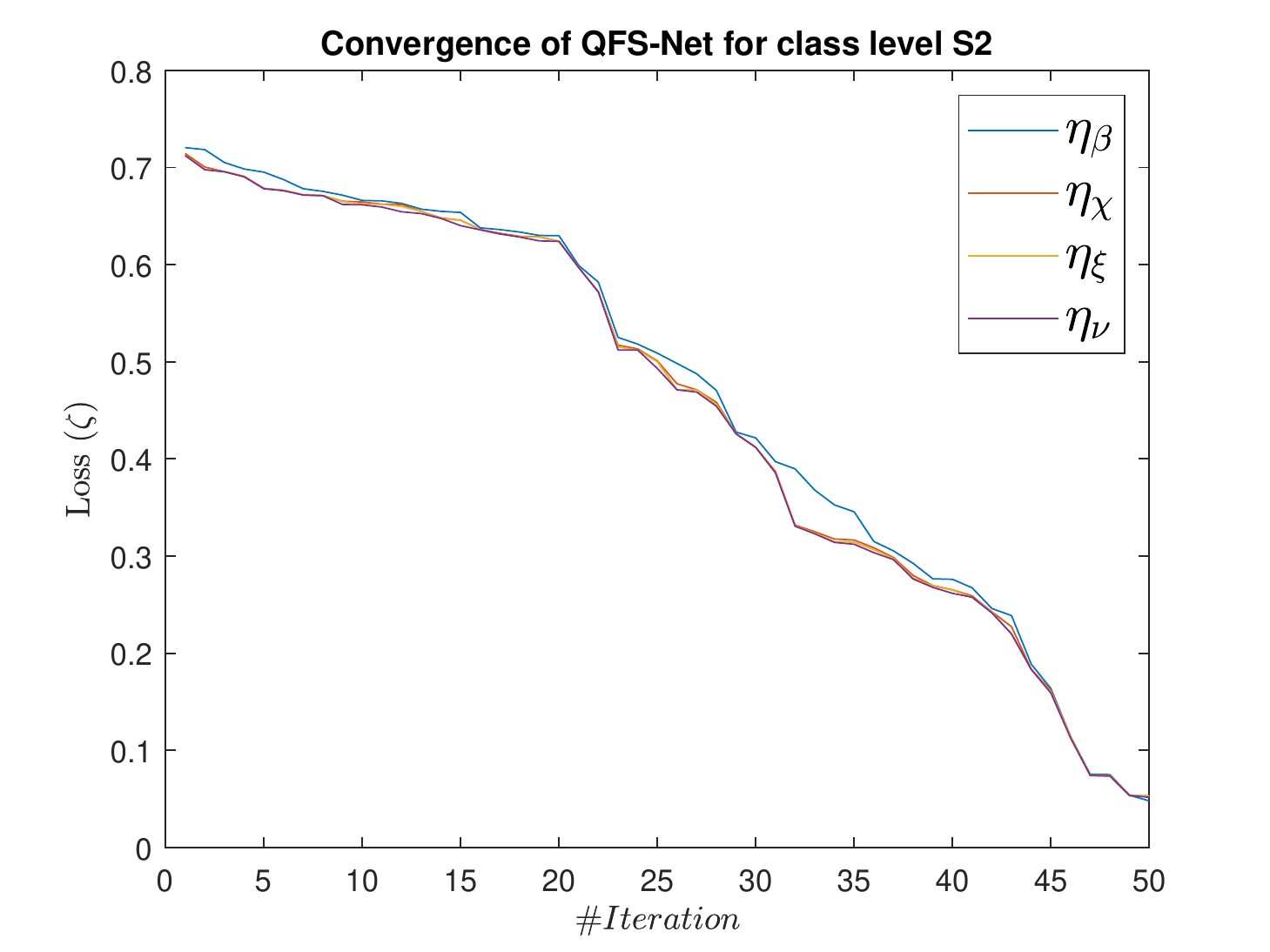}}
 \subcaptionbox{QFS-Net, S3}{\includegraphics[width=1.5in]{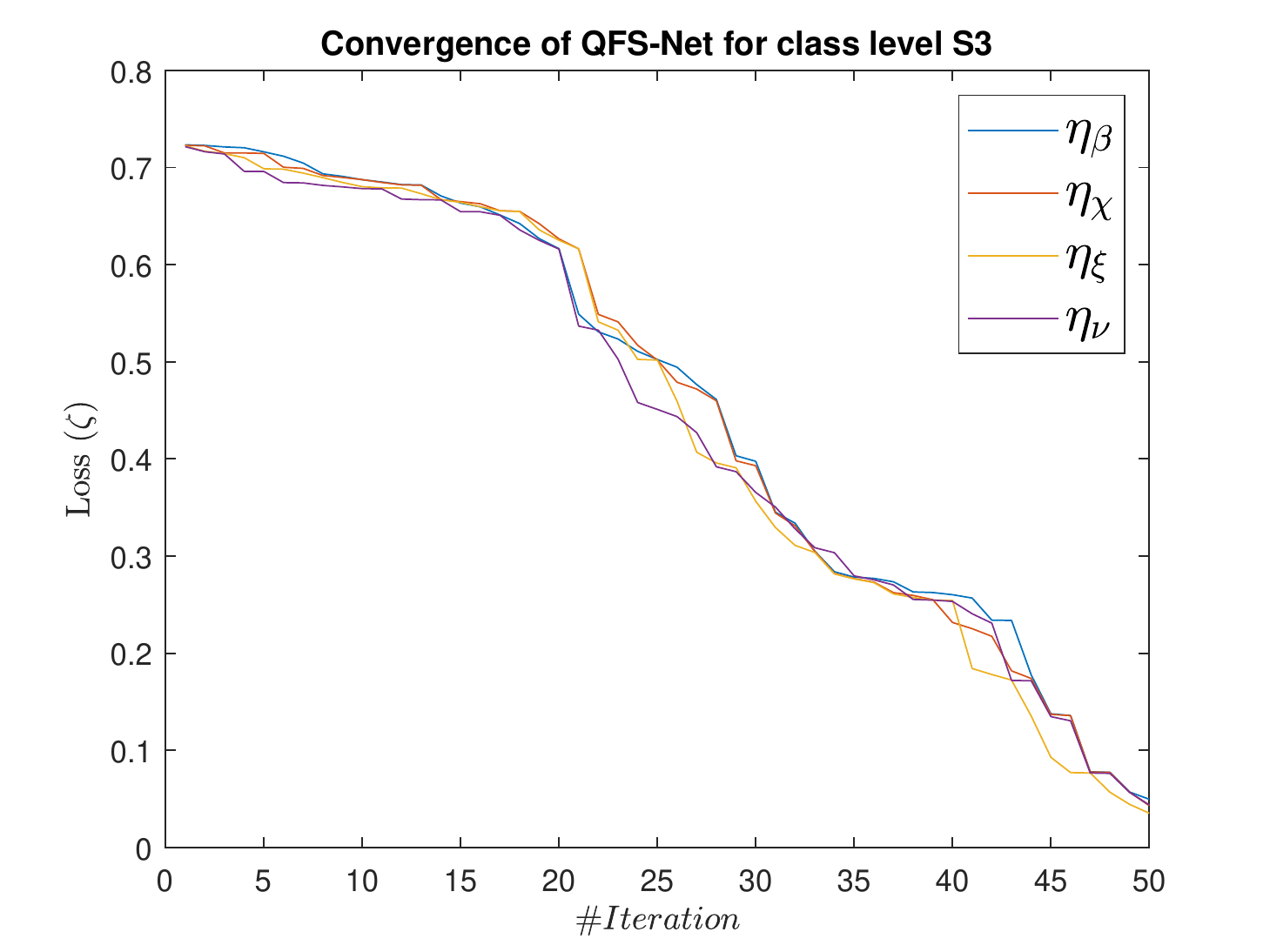}}
 \subcaptionbox{QFS-Net, S4}{\includegraphics[width=1.5in]{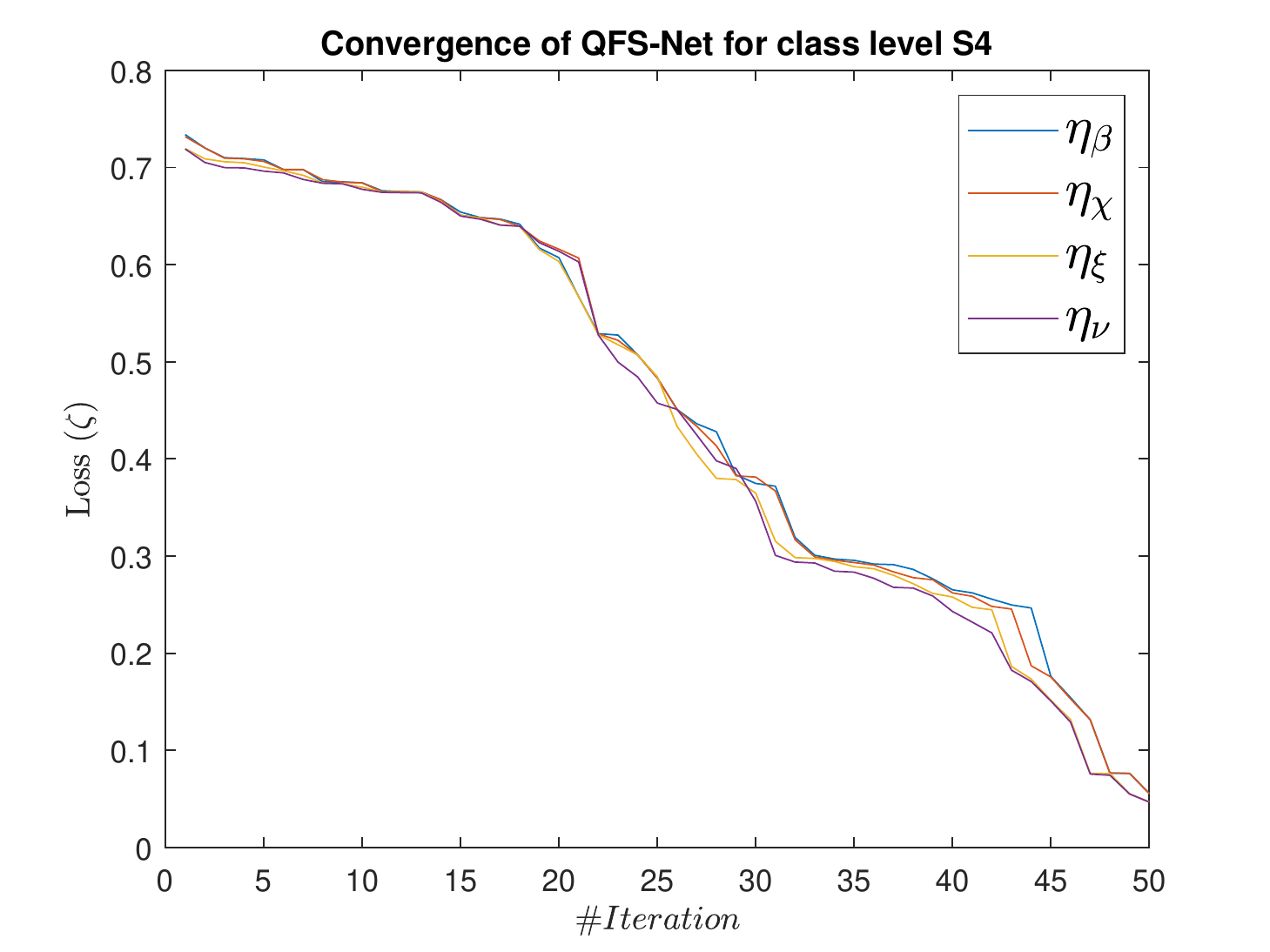}}
	\caption{Convergence analyses of the suggested \emph{qutrit}-inspired QFS-Net and \emph{qubits} embedded QIS-Net~\cite{konar4} for four different activation schemes}
	\label{fig:conv}
\end{figure}

\begin{figure}[htbp]
\centering
 \subcaptionbox{$\eta_{\beta}$}{\includegraphics[width=1.5in]{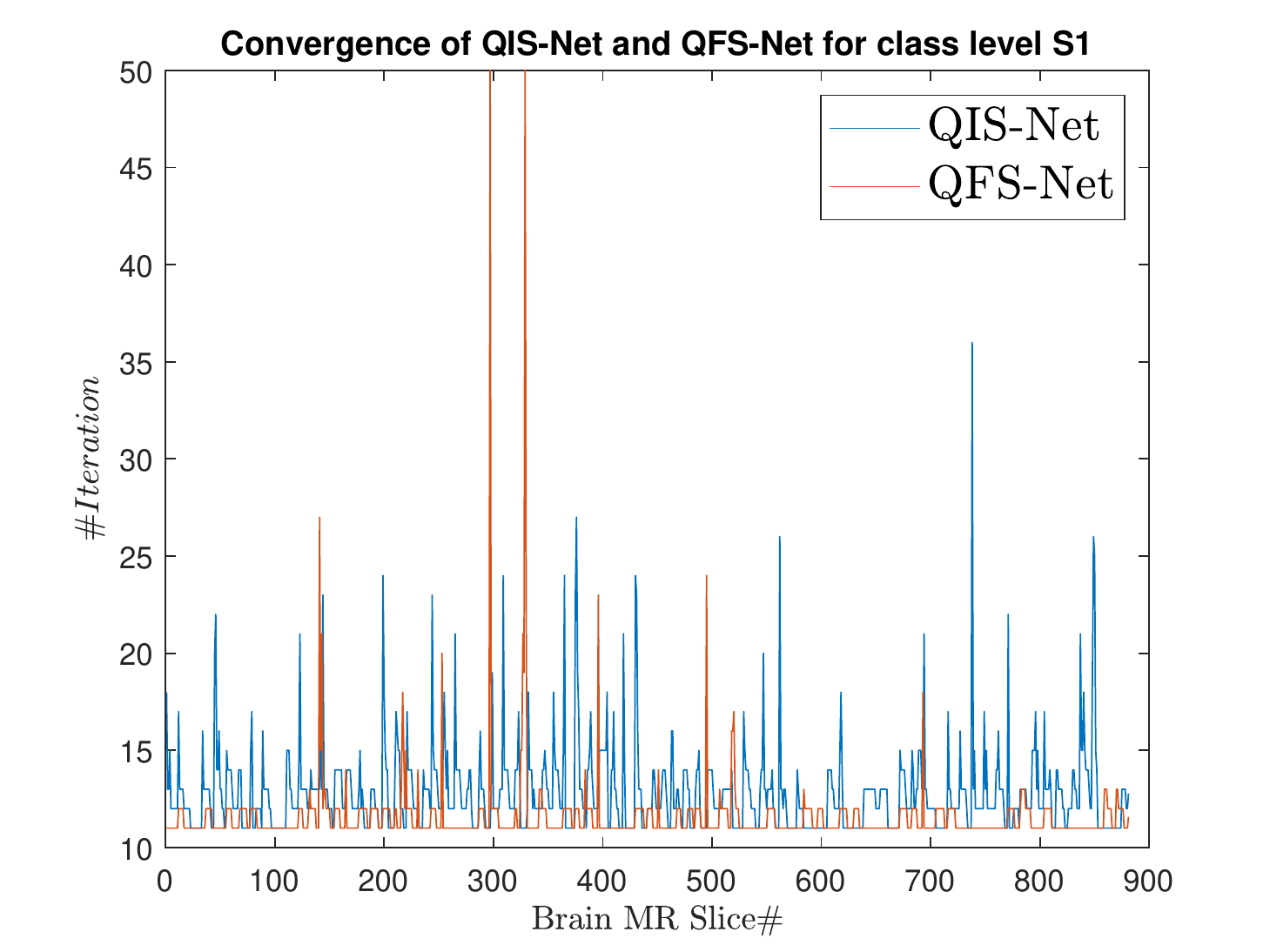}}
 \subcaptionbox{$\eta_{\chi}$}{\includegraphics[width=1.5in]{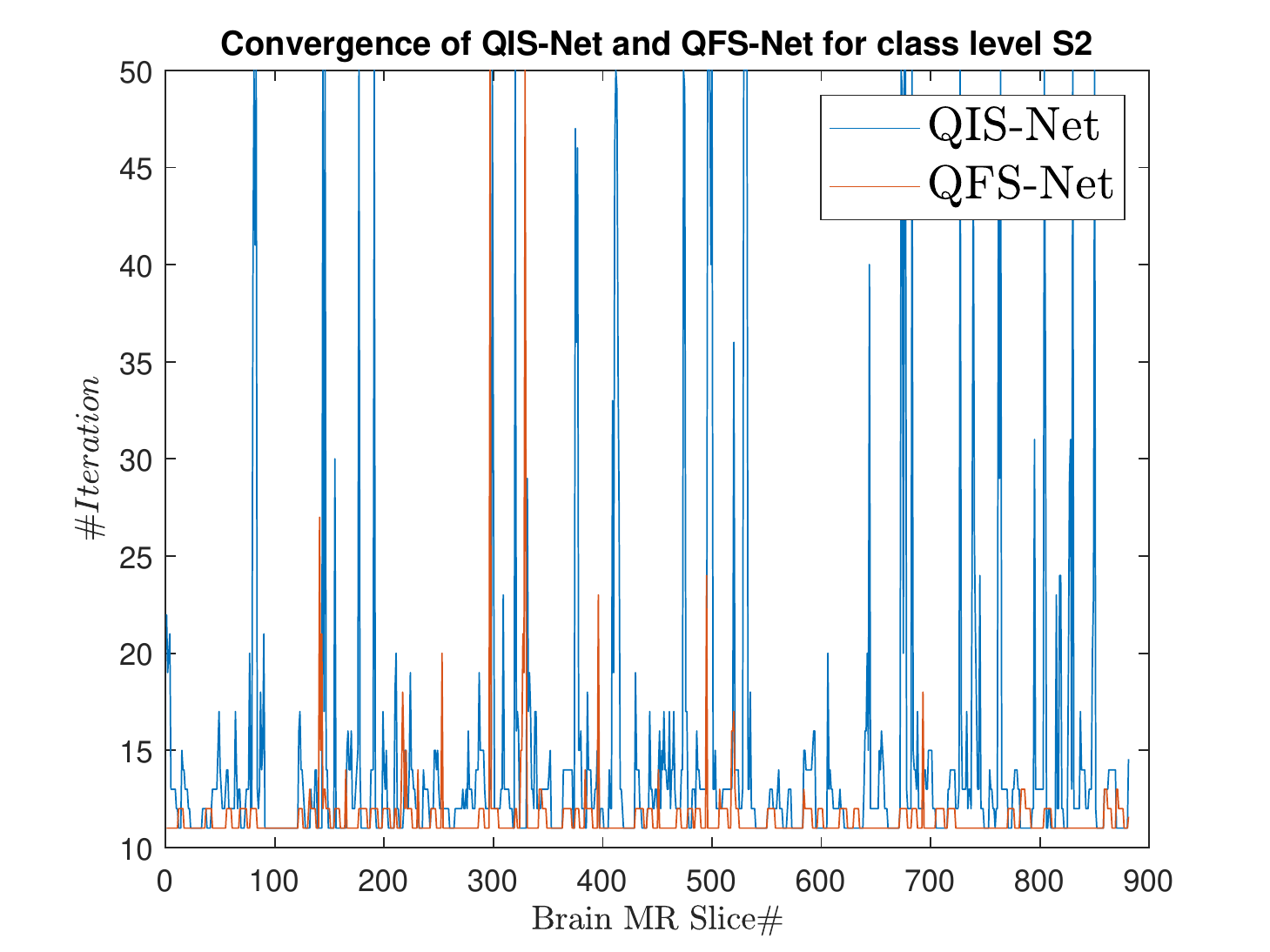}}
 \subcaptionbox{$\eta_{\xi}$}{\includegraphics[width=1.5in]{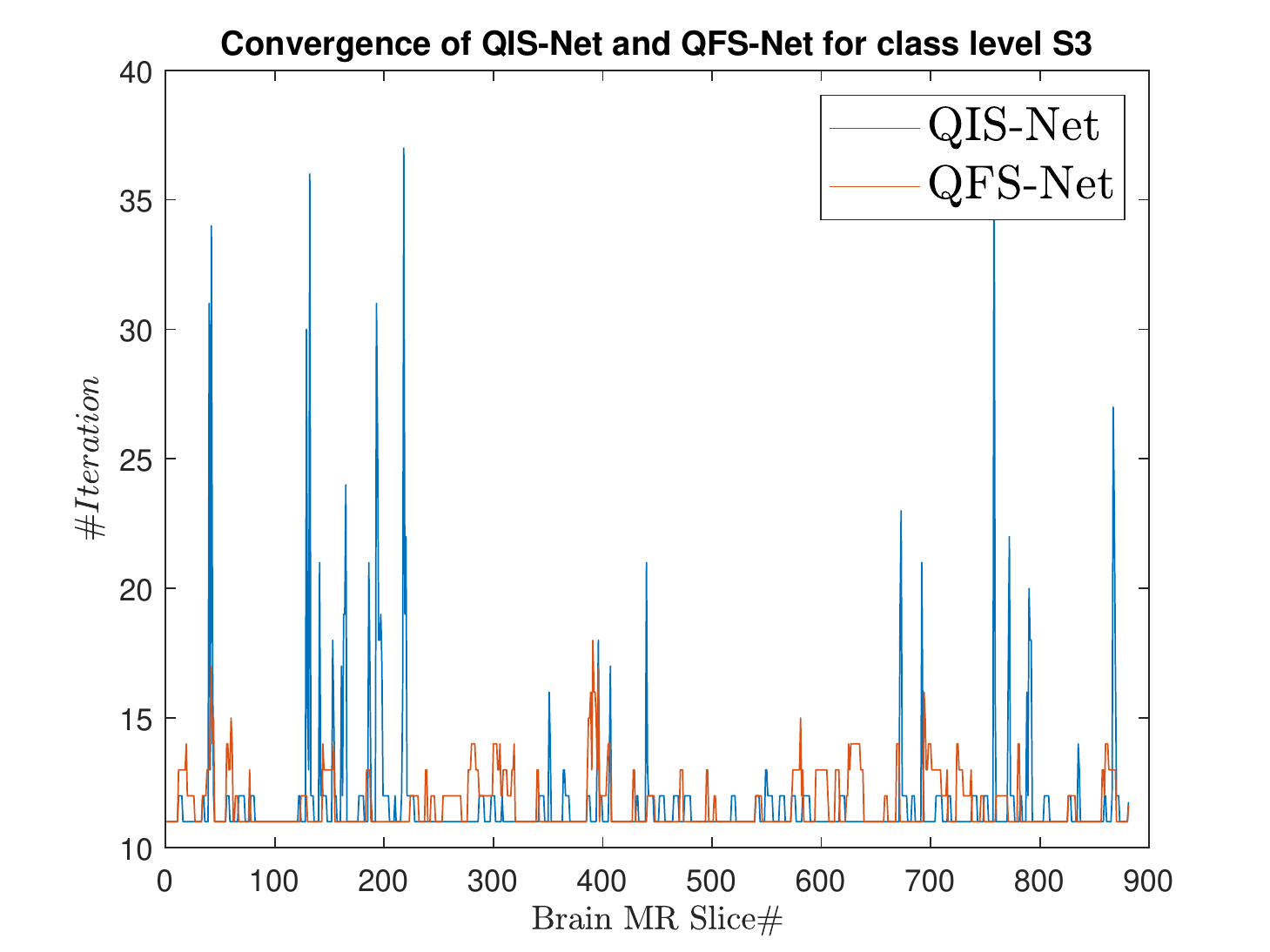}}
 \subcaptionbox{$\eta_{\nu}$}{\includegraphics[width=1.5in]{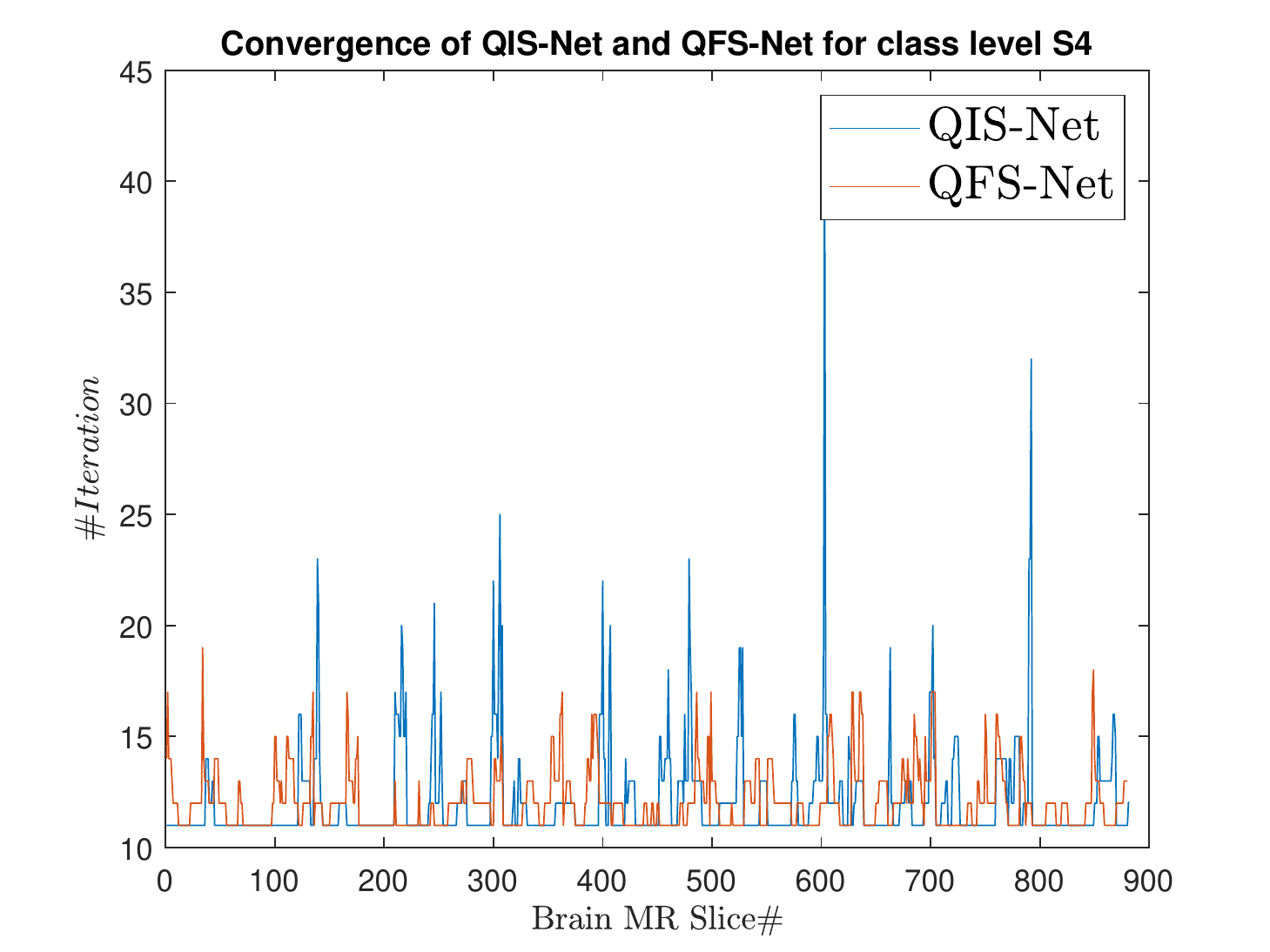}}
\caption{Average number of iterations of each brain slice using QFS-Net based on \emph{qutrit} and QIS-Net~\cite{konar4} based on \emph{qubits} for four various thresholding schemes (a)$\eta_{\beta}$, (b)$\eta_{\chi}$, (c)$\eta_{\xi}$, (d)$\eta_{\nu}$ using class level $S_2$~\cite{konar4}}
\label{fig:iteration}
\end{figure}

The interconnection weights and the activation of QFS-Net architecture are updated using a Hadamard gate ($\mathcal{H}$) working on \emph{qutrit} as follows.
\begin{equation}
\mathcal{H}(|\theta^{\iota+1}\rangle)=\frac{1}{\sqrt{3}}\left[ {{\begin{array}{*{20}c}
		1 & 1 & 1  \hfill \\
		1 & {e^{\jmath \frac{2\pi}{3}\triangle \omega}} & {e^{-\jmath\frac{2\pi}{3}\triangle \omega}}  \hfill \\
		1 & {e^{-\jmath\frac{2\pi}{3}\triangle \omega}} & {e^{\jmath\frac{2\pi}{3}\triangle \omega}} \hfill \\
		\end{array} }} \right]\mathcal{H}(|\theta^{\iota}\rangle)
\end{equation}
\begin{equation}
\mathcal{H}(|\xi^{\iota+1}\rangle)=\frac{1}{\sqrt{3}}\left[ {{\begin{array}{*{20}c}
		1 & 1 & 1  \hfill \\
		1 & {e^{\jmath\frac{2\pi}{3}\triangle \gamma}} & {e^{-\jmath\frac{2\pi}{3}\triangle \gamma}}  \hfill \\
		1 & {e^{-\jmath\frac{2\pi}{3}\triangle \gamma}} & {e^{\jmath\frac{2\pi}{3}\triangle \gamma}} \hfill \\
		\end{array} }} \right]\mathcal{H}(|\xi^{\iota}\rangle)
\end{equation}
where
\begin{equation}
\omega^{\iota+1}=\omega^{\iota}+\triangle \omega^{\iota}
\label{eqn_theta_t}
\end{equation}
and
\begin{equation}
\gamma^{\iota+1}=\gamma^{\iota}+\triangle \gamma^{\iota}
\label{eqn_omega_t}
\end{equation}
The suitable tailoring of the phase angle in the Hadamard gate advocates the stability of the QFS-Net or its convergence which is very crucial for self-supervised networks where the loss function (here error function) is dependent on the interconnection weights. Hence, the phase angles are evaluated using $\triangle \omega^{\iota}$ and $\triangle \gamma^{\iota}$ as given in Equations~\ref{eq:alpha} and \ref{eq:gamma}, respectively. It is worth noting that the \emph{qutrit} based quantum neural network provides faster convergence compared to the classical neural networks. This is due to the fact that whereas the classical neural networks are formed using the multiplication of input vector and the weight vector guided by an activation function, the quantum-based networks incorporate the frequency components of the weights and their inputs thereby enabling faster convergence of the network states. This inherent novel feature of the quantum neural networks facilitates the \emph{qutrit} based fully self-organized quantum algorithm to be employed in QFS-Net to converge super-linearly, as shown in Figure~\ref{fig:conv}. The loss function cum QFS-Net network error function is defined on quantum measurement in the following way. 
\begin{equation}
\zeta (\omega,\gamma)=\frac{1}{N}\sum_{i}^{N}\sum_{k=1}^8 \left[\Theta_{ik}(\omega_{ik},\gamma_i) ^{\iota+1}-\Theta_{ik}(\omega_{ik}, \gamma_i)^{\iota}\right]^2
\end{equation}
where, $\Theta_{ik}(\omega_{ik},\gamma_i)^{\iota}$ represents the true interconnection weight terms of the inter-connection weights $|\theta_{ij}^{\iota} \rangle$ as expressed using the Hadamard gate ($\mathcal{H}$) at an instance ($\iota$). $\zeta (\omega,\gamma)$ is a coherent error function of $\omega$ and $\gamma$. Convergence analysis of the proposed \emph{qutrit}-inspired QFS-Net is provided in Appendix Section~\ref{conv:QFS-Net} and demonstrated experimentally with \emph{qubit} embedded QIS-Net~\cite{konar4} as shown in Figure~\ref{fig:conv}. It can be summarized that the convergence of the QFS-Net is faster than that of the QIS-Net and also follows super-linearity. This claim is also substantiated by the number of iterations required to converge for each image slice in QFS-Net and QIS-Net as illustrated in Figure~\ref{fig:iteration}.


\section{Results and Discussion}
\label{results:disscuss}

\subsection{Data Set}
\label{results:dataset}
Cancer Imaging Archive (TCIA) data are available from the Nature repository~\cite{data} and the experiments have been performed on the same data sets using the suggested QFS-Net model characterized by \emph{qutrits} and an adaptive multi-class quantum sigmoidal (\emph{QSig}) activation function. In contrast with the automatic brain lesion segmentation, four distinct activation schemes have been tested, and experiments are also performed using Quantum-Inspired Self-supervised Network (QIS-Net)~\cite{konar4}, U-Net~\cite{olaf} and URes-Net~\cite{guerrero} architectures. The U-Net~\cite{olaf} and URes-Net~\cite{guerrero} architectures are trained with $2000$ MR images and validated and tested on $120$ and $880$ contrast-enhanced Dynamic Susceptibility Contrast (DSC) MR images, respectively. The QIS-Net and the proposed QFS-Net are also tested on the same number of $880$ contrast-enhanced DSC MR images. 


\subsection{Experimental Setup}
\label{experiment:result}

In this current work, extensive experiments have been carried out on $3000$ Dynamic Susceptibility Contrast (DSC) brain MR images of Glioma patients from TCIA data sets of size $512 \times 512$ using Nvidia RTX 2070 GPU System with high-performance systems with MATLAB 2020a and Python 3.6. The 2D segmented images are processed through a 2D binary circular mask to obtain the brain lesion in the suggested QFS-Net framework. The lesion or brain tumor detection mask is binarized using a threshold of $0.5$, and in the case of QFS-Net and QIS-Net~\cite{konar4}, it is seen that with a radius of $5$ pixels, the segmented ROIs perform optimally while compared with the human expert segmented images. Experiments are also performed on two recently developed CNN architectures suitable for medical image segmentation viz., convolutional U-Net~\cite{olaf} and Residual U-Net (URes-Net)~\cite{guerrero} available in GitHub. The U-Net and URes-Net networks are rigorously trained using the stochastic gradient descent algorithm with learning rate $0.001$ and batch size $32$ allowing maximum $50$ epochs to converge. The segmented output images resemble in size with the dimensions of the binary mask and the outcome $1$ is considered as tumor region and $0$ as background in detecting complete tumor. The pixel by pixel comparison with the manually segmented regions of interest or lesion mask allows evaluating the dice similarity, which is considered as a standard evaluation procedure in automatic medical image segmentation. The evaluation process involves the manually segmented lesion mask as ground truth, and each 2D pixel is predicted as either True Positive ($T_{RP}$) or True Negative ($T_{RN}$) or False Positive ($T_{RN}$) or False Negative ($F_{LN}$).     \\
The suggested \emph{qutrit}-inspired fully self-supervised shallow quantum learning model is experimented with the multi-level gray-scale images using distinct classes $L=4,5, 6, 7$ and $8$ characterized by an adaptive multi-class quantum sigmoidal (\emph{QSig}) activation function. In this experiment, the steepness in the \emph{QSig} activation, $\lambda$ is varied in the range $0.23$ to $0.24$ with step size $0.001$. It has been observed that in majority cases, $\lambda = 0.239$ yields optimal performance. The empirical goodness measures [Positive Predictive Value ($PPV$), Sensitivity ($SS$), Accuracy ($ACC$) and Dice Similarity($DS$)~\cite{alex}] are assessed to evaluate the experimental outcome using four thresholding schemes $(\eta_\beta, \eta_\chi, \eta_\xi,\eta_\nu)$~\cite{konar4,bhatt4} as discussed in the supplementary materials section for different level sets. The dice score is often used to measure the similarity of the segmented brain lesions and regions of interest (ROIs). 

\subsection{Experimental Results}
\label{result:exp}

\begin{figure}[htbp]
\centering
 \subcaptionbox{Input Slice $\# 37$}{\includegraphics[width=0.75in]{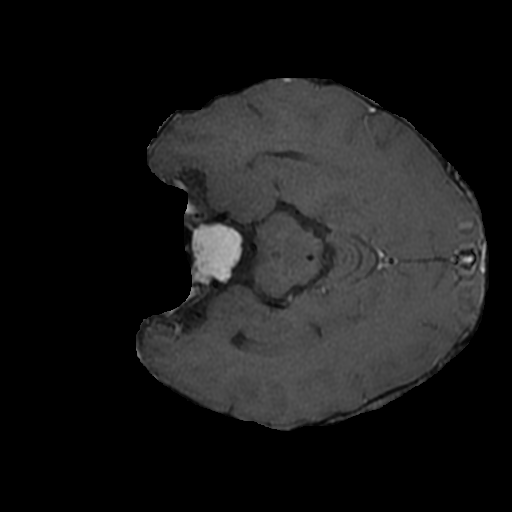}}
 \subcaptionbox{Input Slice $\# 69$}{\includegraphics[width=0.75in]{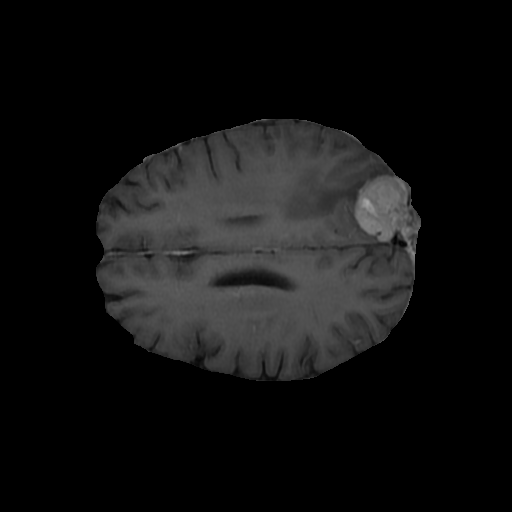}}
 \subcaptionbox{ROI Slice $\# 37$}{\includegraphics[width=0.75in]{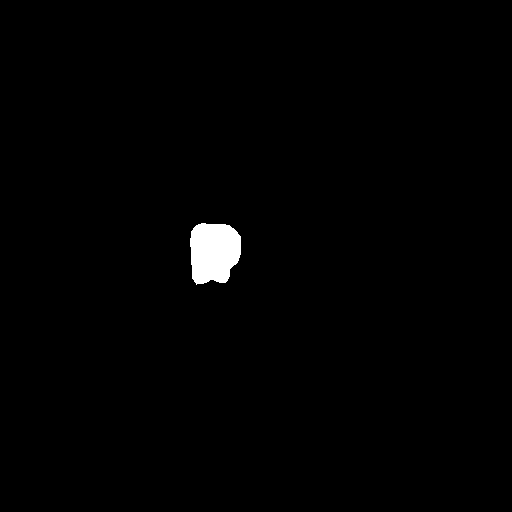}}
 \subcaptionbox{ROI Slice $\# 69$}{\includegraphics[width=0.75in]{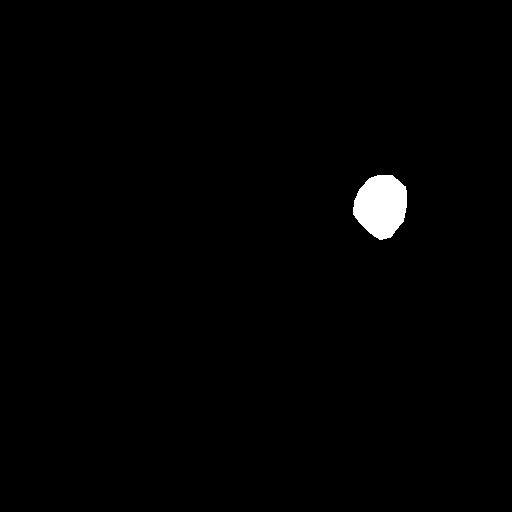}}
\caption{Dynamic Susceptibility Contrast (DSC) skull stripped brain MR images with size $512 \times 512$ and manually segmented ROI slices~\cite{data}}
\label{fig:Input}
\end{figure}

\begin{figure}[htp]
	\centering
 \subcaptionbox{$\eta_{\beta}$}{\includegraphics[width=0.75in]{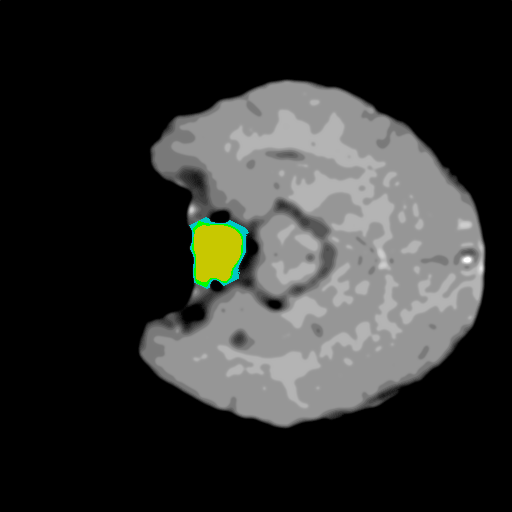}}
 \subcaptionbox{$\eta_{\chi}$}{\includegraphics[width=0.75in]{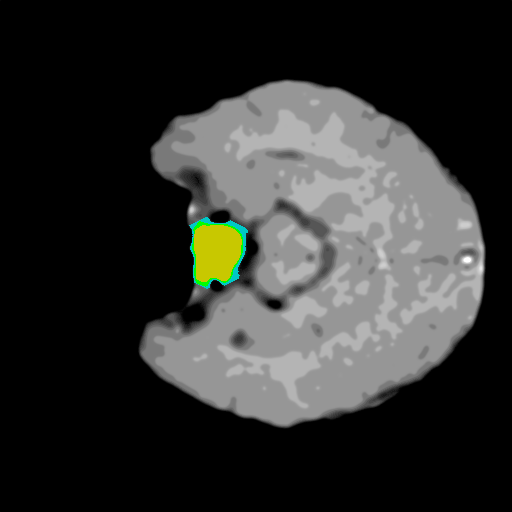}}
 \subcaptionbox{$\eta_{\xi}$}{\includegraphics[width=0.75in]{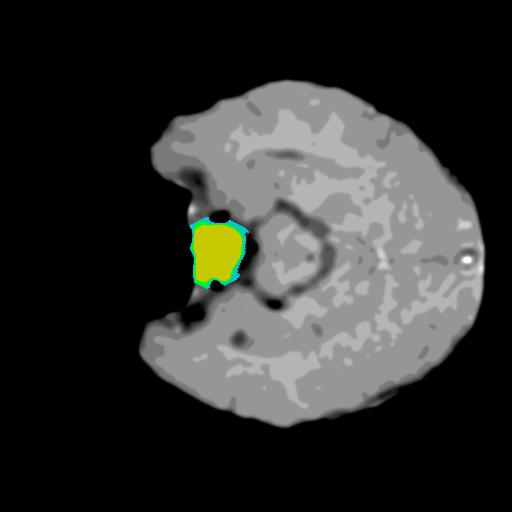}}
 \subcaptionbox{$\eta_{\nu}$}{\includegraphics[width=0.75in]{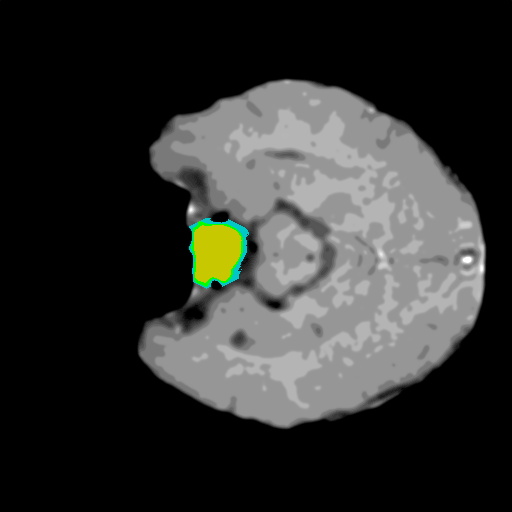}}
 \subcaptionbox{$\eta_{\beta}$}{\includegraphics[width=0.75in]{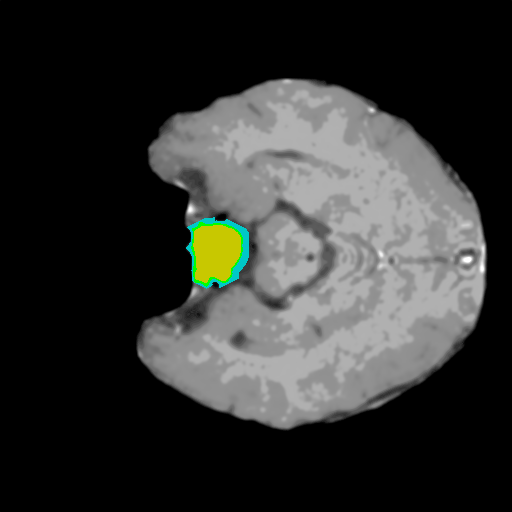}}
 \subcaptionbox{$\eta_{\chi}$}{\includegraphics[width=0.75in]{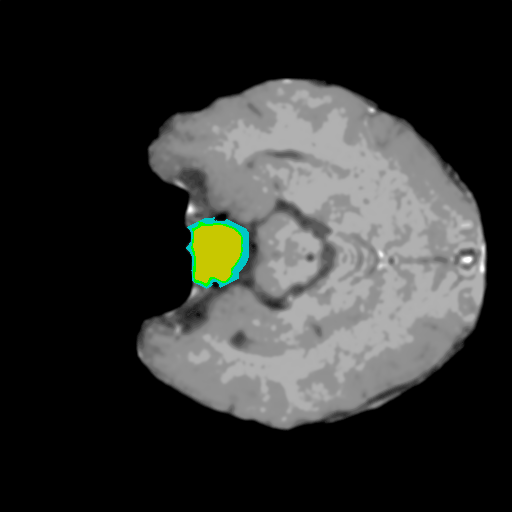}}
 \subcaptionbox{$\eta_{\xi}$}{\includegraphics[width=0.75in]{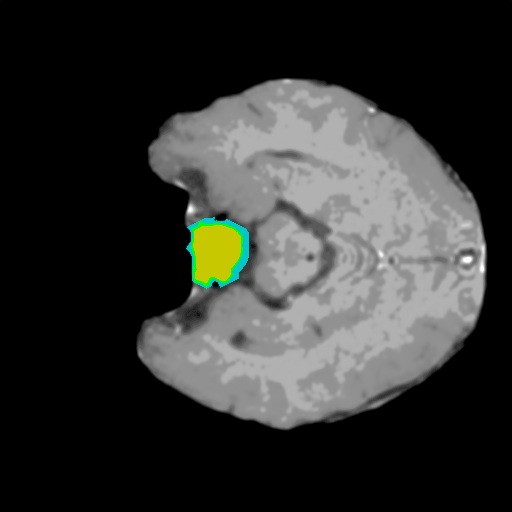}}
 \subcaptionbox{$\eta_{\nu}$}{\includegraphics[width=0.75in]{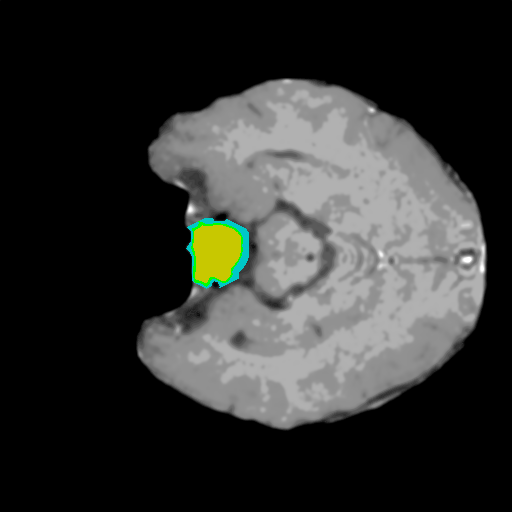}}
 \subcaptionbox{$\eta_{\beta}$}{\includegraphics[width=0.75in]{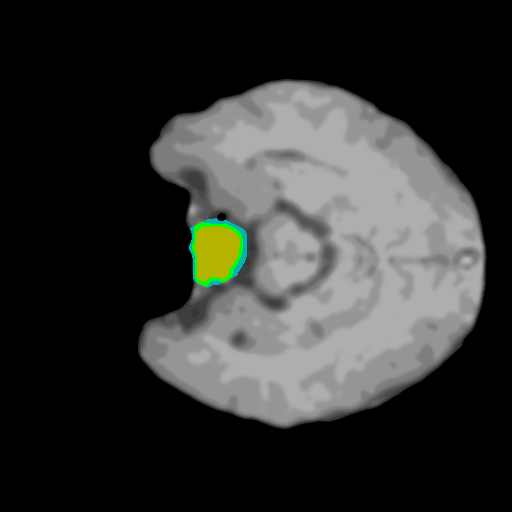}}
 \subcaptionbox{$\eta_{\chi}$}{\includegraphics[width=0.75in]{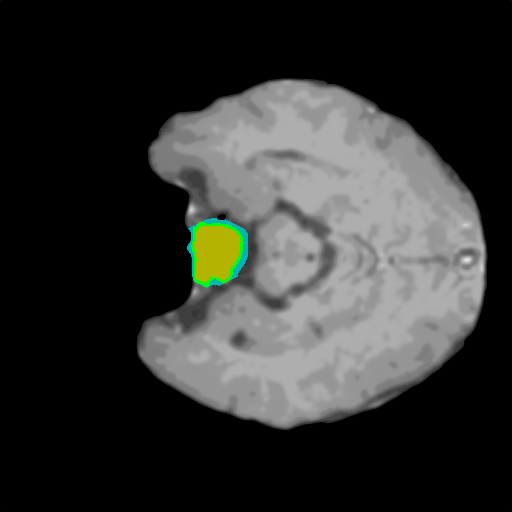}}
 \subcaptionbox{$\eta_{\xi}$}{\includegraphics[width=0.75in]{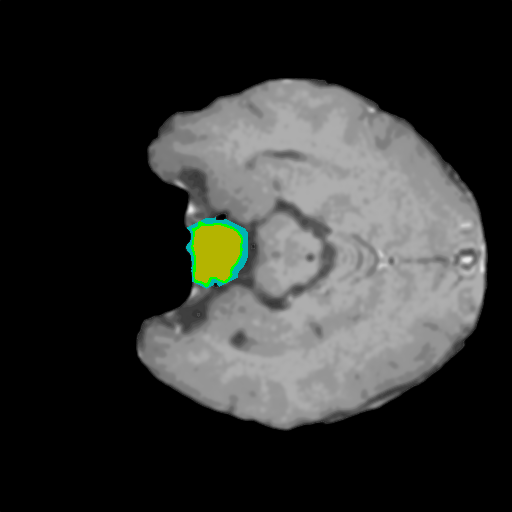}}
 \subcaptionbox{$\eta_{\nu}$}{\includegraphics[width=0.75in]{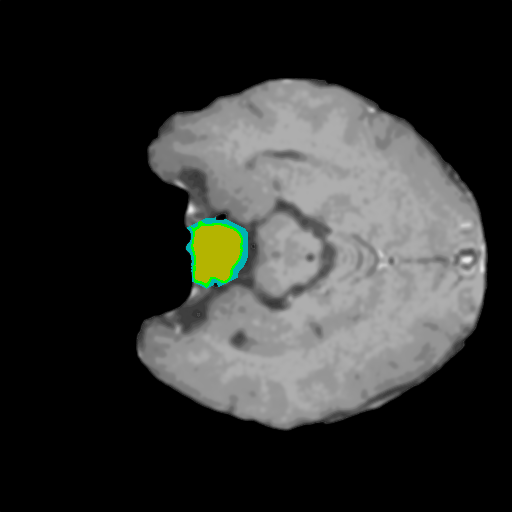}}
 \subcaptionbox{$\eta_{\beta}$}{\includegraphics[width=0.75in]{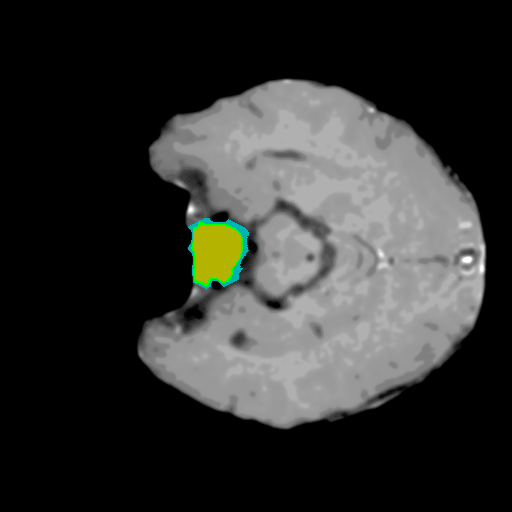}}
 \subcaptionbox{$\eta_{\chi}$}{\includegraphics[width=0.75in]{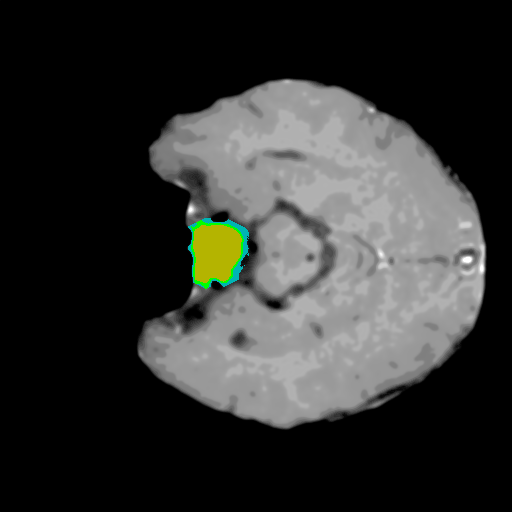}}
 \subcaptionbox{$\eta_{\xi}$}{\includegraphics[width=0.75in]{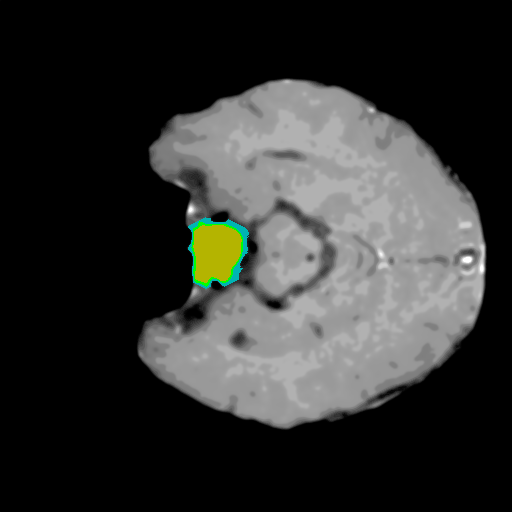}}
 \subcaptionbox{$\eta_{\nu}$}{\includegraphics[width=0.75in]{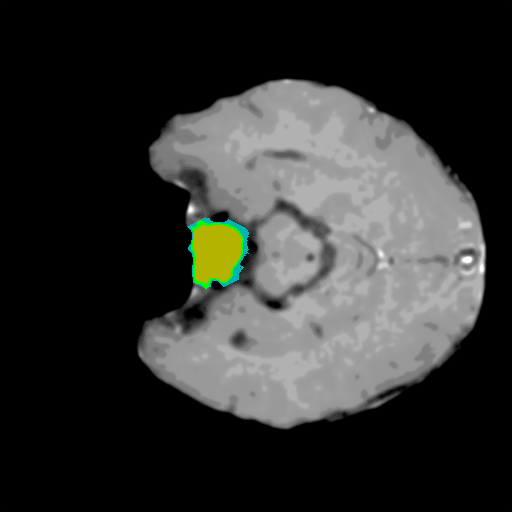}}
  \caption{Demonstration of QFS-Net segmented images followed by essential post-processed outcome on the slice no. $37$~\cite{data} for class level $L=8$ with four distinct activation schemes ($\eta_{\beta}, \eta_{\chi}, \eta_{\xi}, \eta_{\nu}$) with class-levels $(a-d)$ for $S_1$ , $(e-h)$ for $S_2$, $(i-l)$ for $S_3$,and $(m-p)$ for $S_4$~\cite{konar4}}
  \label{fig:C8-37}
\end{figure}

\begin{figure}[htp]
	\centering
 \subcaptionbox{$\eta_{\beta}$}{\includegraphics[width=0.75in]{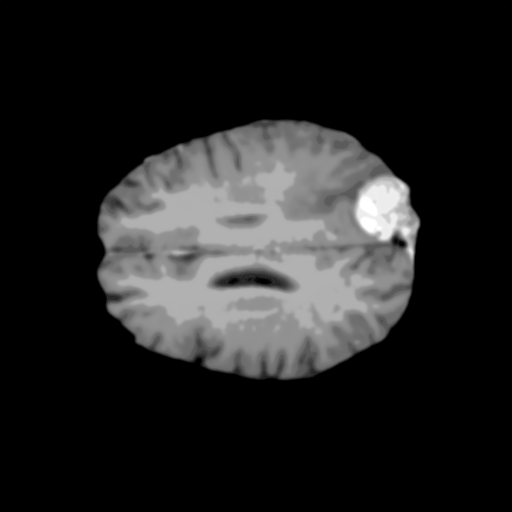}}
 \subcaptionbox{$\eta_{\chi}$}{\includegraphics[width=0.75in]{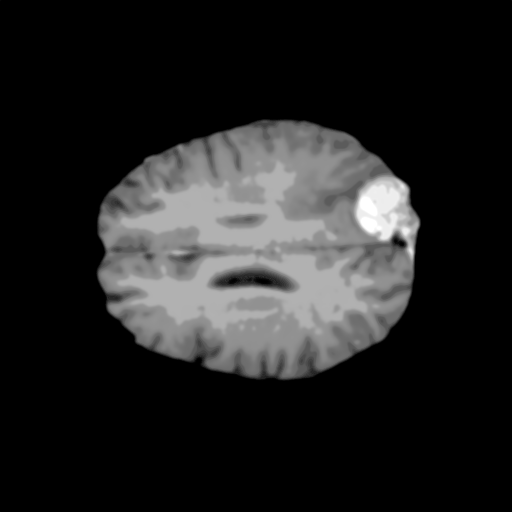}}
 \subcaptionbox{$\eta_{\xi}$}{\includegraphics[width=0.75in]{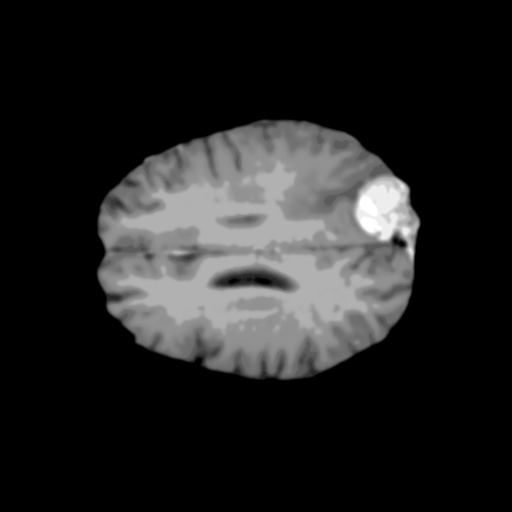}}
 \subcaptionbox{$\eta_{\nu}$}{\includegraphics[width=0.75in]{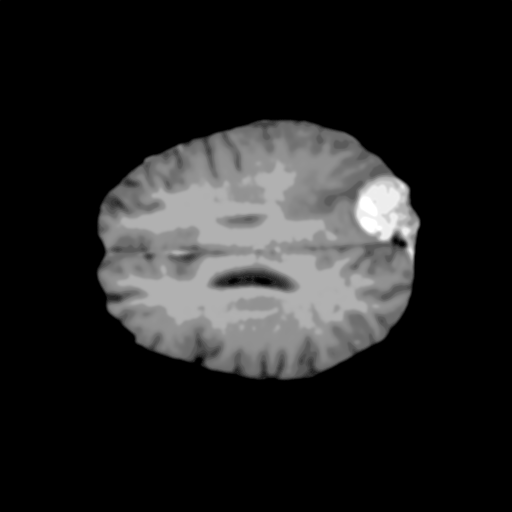}}
  \subcaptionbox{$\eta_{\beta}$}{\includegraphics[width=0.75in]{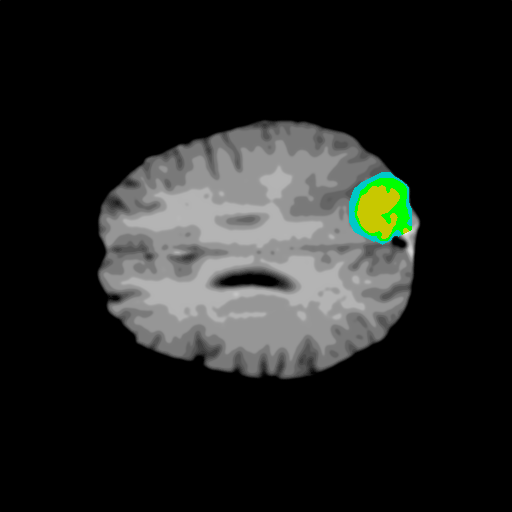}}
 \subcaptionbox{$\eta_{\chi}$}{\includegraphics[width=0.75in]{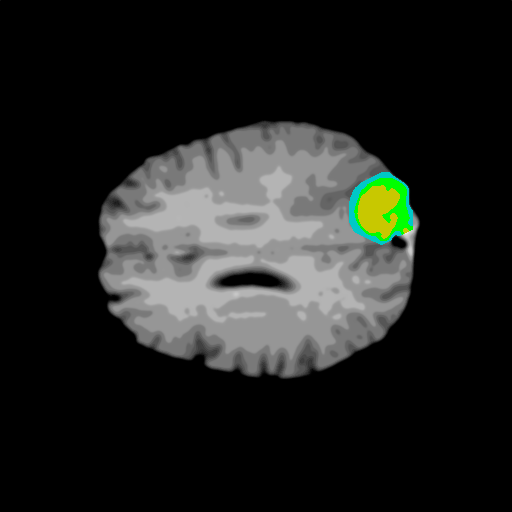}}
 \subcaptionbox{$\eta_{\xi}$}{\includegraphics[width=0.75in]{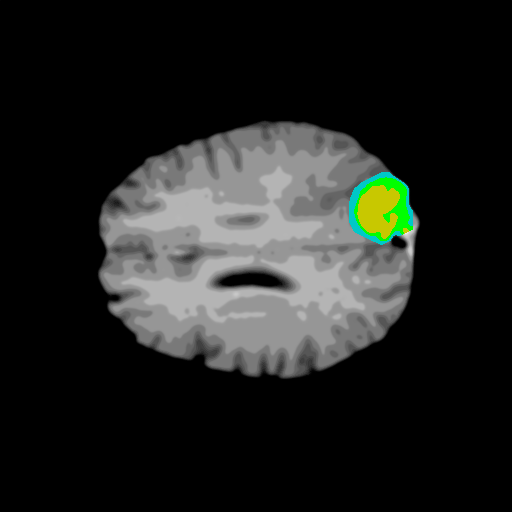}}
 \subcaptionbox{$\eta_{\nu}$}{\includegraphics[width=0.75in]{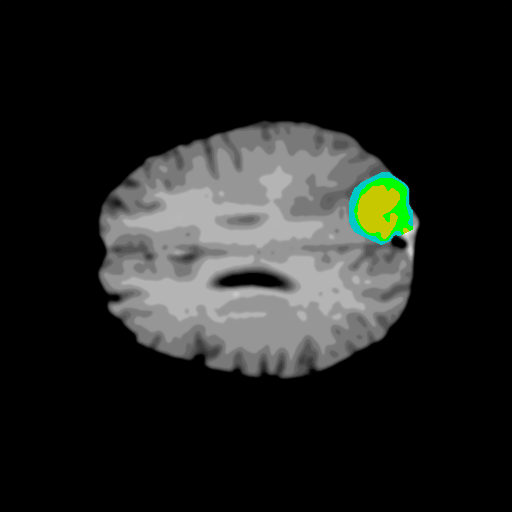}}
  \subcaptionbox{$\eta_{\beta}$}{\includegraphics[width=0.75in]{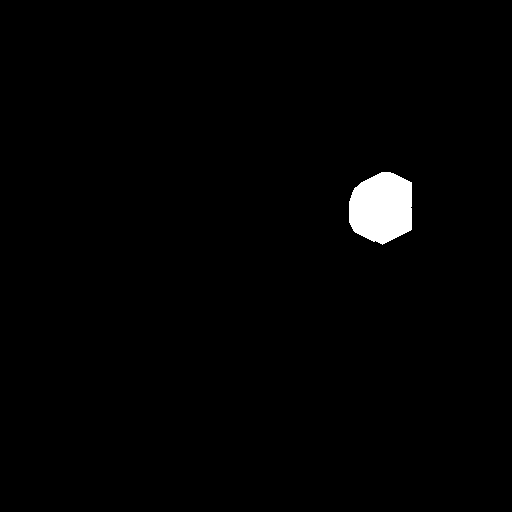}}
 \subcaptionbox{$\eta_{\chi}$}{\includegraphics[width=0.75in]{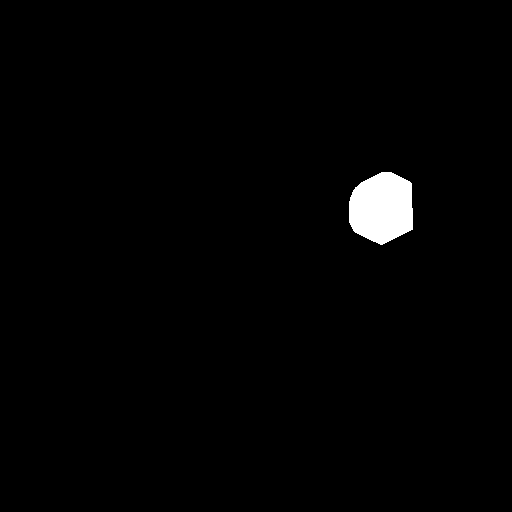}}
 \subcaptionbox{$\eta_{\xi}$}{\includegraphics[width=0.75in]{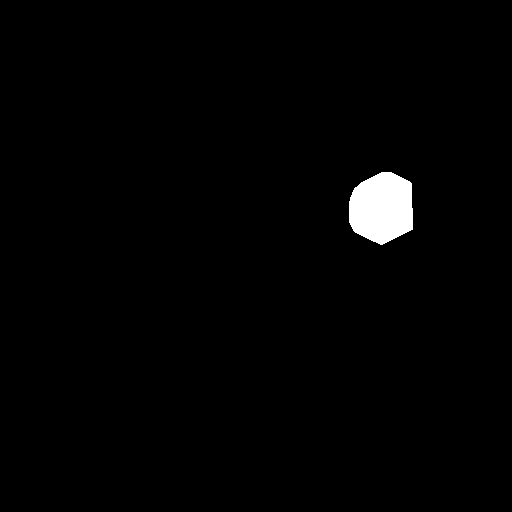}}
 \subcaptionbox{$\eta_{\nu}$}{\includegraphics[width=0.75in]{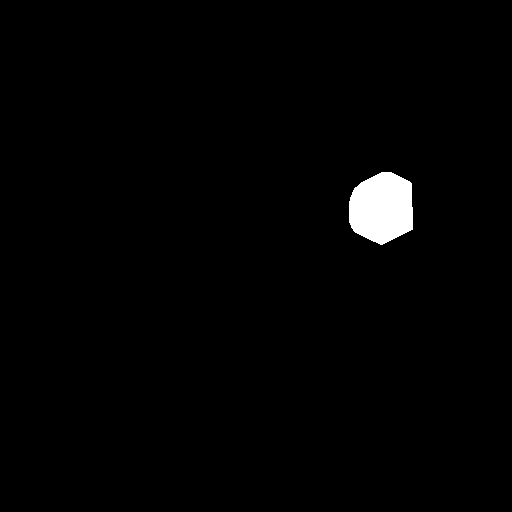}}
 \caption{Segmented ROIs describing the complete tumor region after the post-processing using the proposed QFS-Net on slice $\# 69$~\cite{data} using $L=8$ transition levels with four different thresholding schemes ($\eta_{\beta}, \eta_{\chi}, \eta_{\xi}, \eta_{\nu}$) $(a-e)$ with class-level $S1$~\cite{konar4}}
  \label{fig:C8-S1}
\end{figure}

\begin{figure}[htbp]
	\centering
 \subcaptionbox{$\eta_{\beta}$}{\includegraphics[width=0.75in]{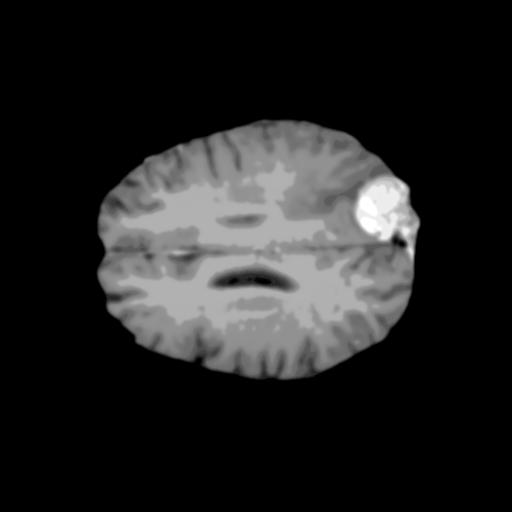}}
 \subcaptionbox{$\eta_{\chi}$}{\includegraphics[width=0.75in]{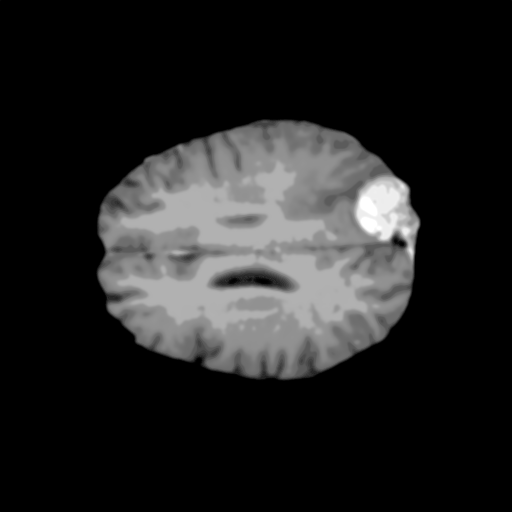}}
 \subcaptionbox{$\eta_{\xi}$}{\includegraphics[width=0.75in]{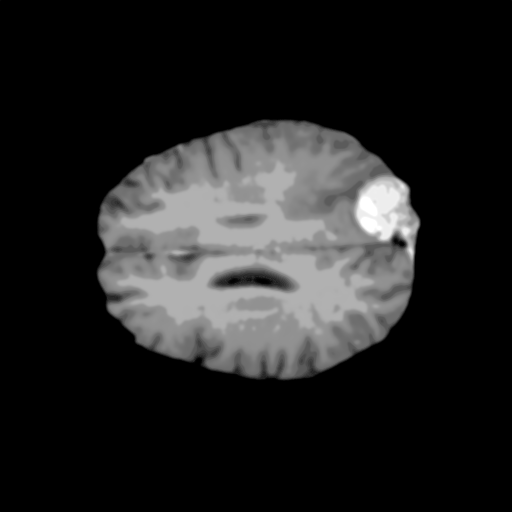}}
 \subcaptionbox{$\eta_{\nu}$}{\includegraphics[width=0.75in]{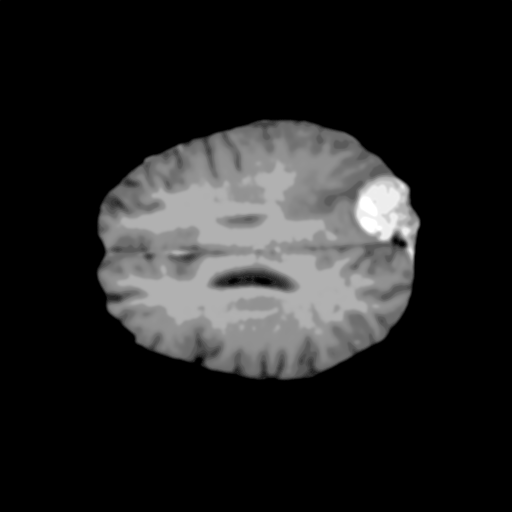}}
 \subcaptionbox{$\eta_{\beta}$}{\includegraphics[width=0.75in]{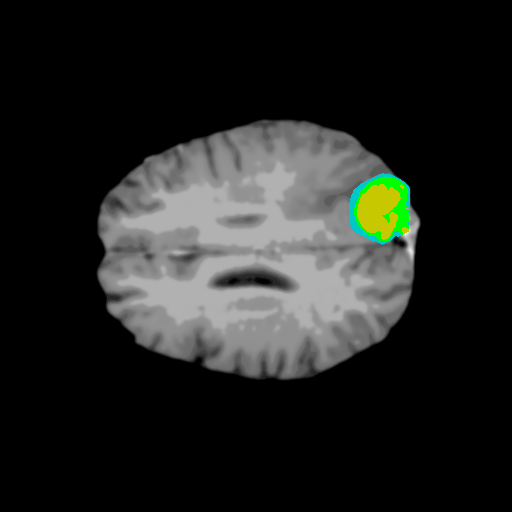}}
 \subcaptionbox{$\eta_{\chi}$}{\includegraphics[width=0.75in]{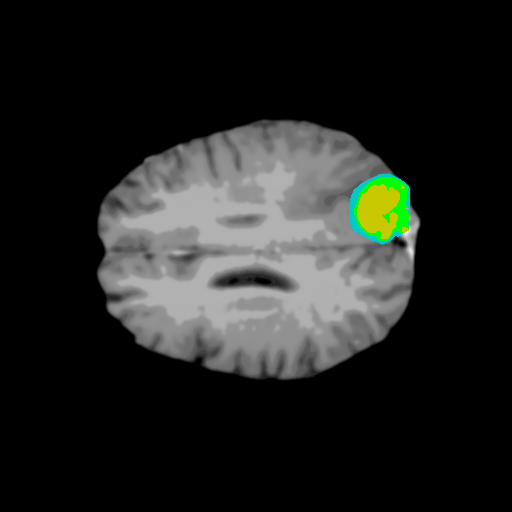}}
 \subcaptionbox{$\eta_{\xi}$}{\includegraphics[width=0.75in]{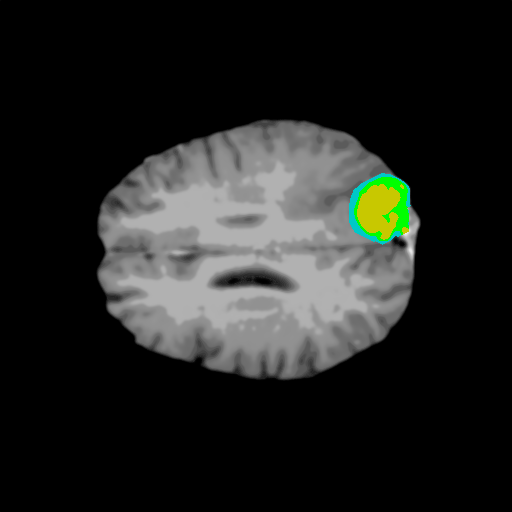}}
 \subcaptionbox{$\eta_{\nu}$}{\includegraphics[width=0.75in]{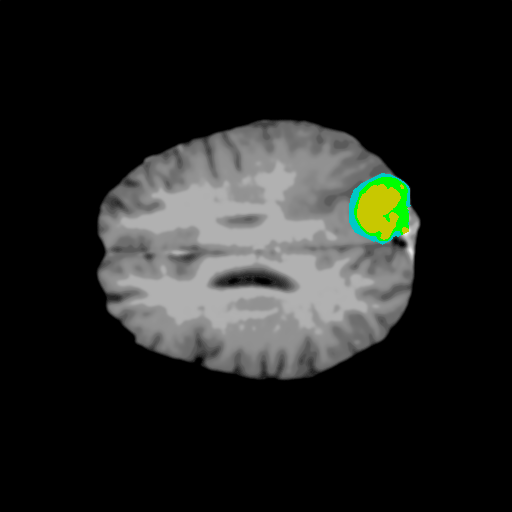}}
 \subcaptionbox{$\eta_{\beta}$}{\includegraphics[width=0.75in]{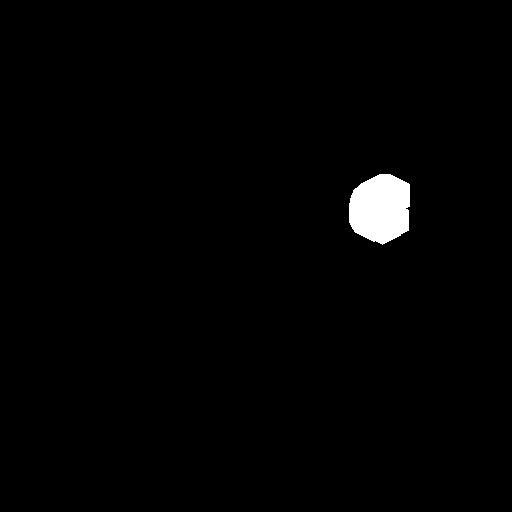}}
 \subcaptionbox{$\eta_{\chi}$}{\includegraphics[width=0.75in]{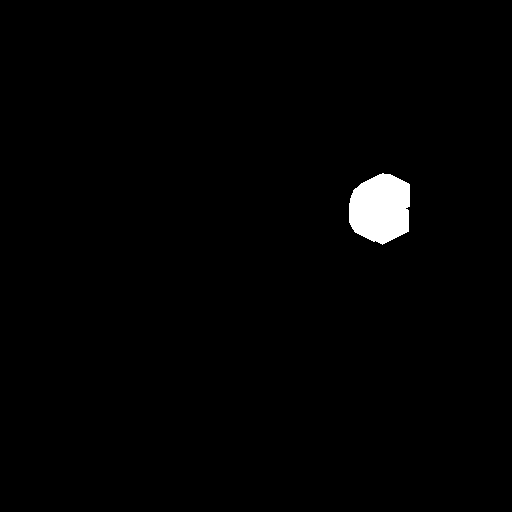}}
 \subcaptionbox{$\eta_{\xi}$}{\includegraphics[width=0.75in]{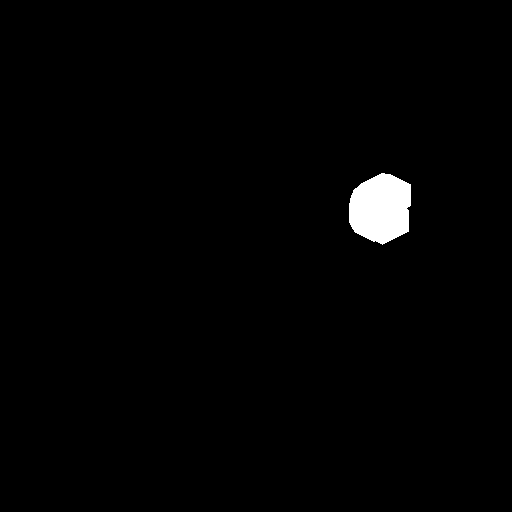}}
 \subcaptionbox{$\eta_{\nu}$}{\includegraphics[width=0.75in]{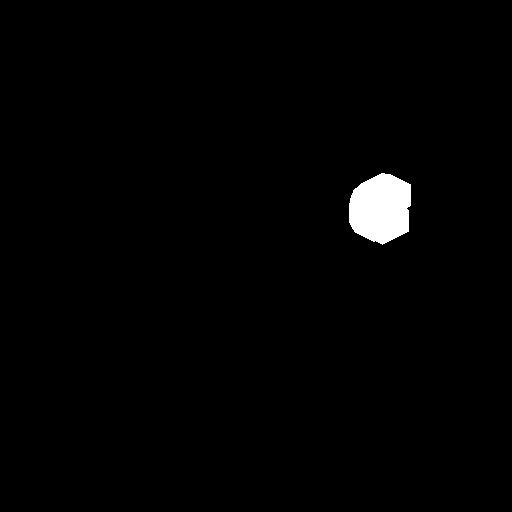}}
	\caption{Segmented ROIs describing the complete tumor region after the post-processing using the proposed QFS-Net on slice $\# 69$~\cite{data} using $L=8$ transition levels with four different thresholding schemes ($\eta_{\beta}, \eta_{\chi}, \eta_{\xi}, \eta_{\nu}$) $(a-e)$ with class-level $S_2$~\cite{konar4}}
	\label{fig:C8-S2}
\end{figure}

\begin{figure}[htbp]
	\centering
  \subcaptionbox{QIS-Net}{\includegraphics[width=0.75in]{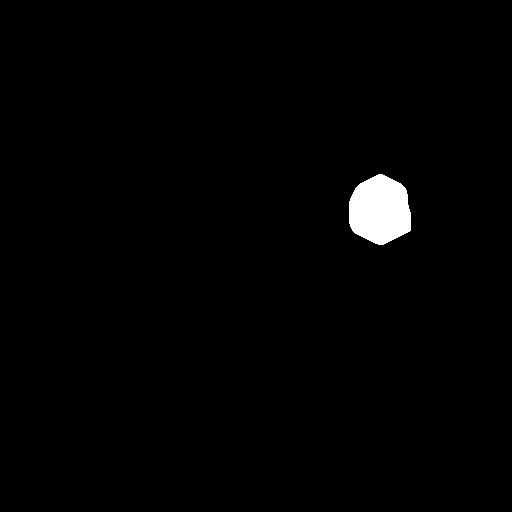}}
 \subcaptionbox{U-Net}{\includegraphics[width=0.75in]{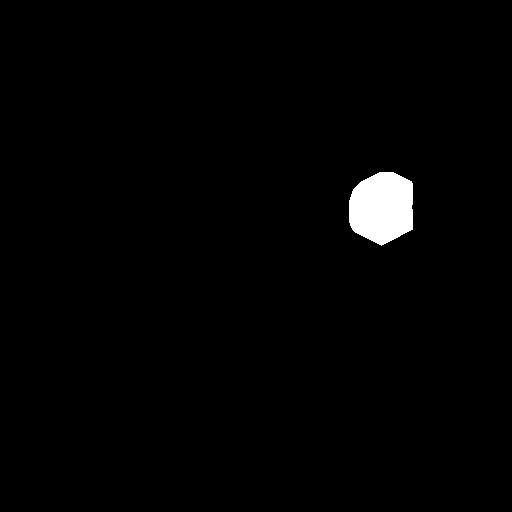}}
 \subcaptionbox{URes-Net}{\includegraphics[width=0.75in]{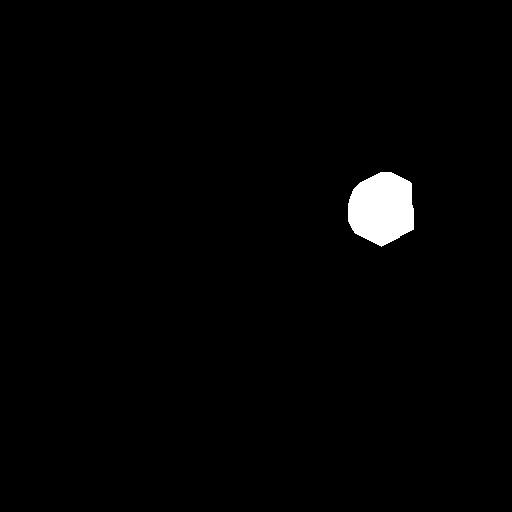}}
	\caption{ROI segmented output slice $\# 69$~\cite{data} masking followed by post processing using $(a)$ QIS-Net~\cite{konar4} ($b$) U-Net~\cite{olaf} ($c$) URes-Net~\cite{guerrero}}
	\label{fig:CNN}
\end{figure}

Extensive experiments have been performed in the current setup, and experimental outcomes are reported with the demonstration of numerical and statistical analyses using the proposed QFS-Net, QIS-Net~\cite{konar4}, convolutional U-Net~\cite{olaf} and Residual U-Net (URes-Net) architectures~\cite{guerrero}. The human expert segmented skull-tripped contrast enhanced DSC brain MR input image slices of size $512 \times 512$ and ROIs are provided in Figure~\ref{fig:Input} as samples. The demonstration of QFS-Net segmented images followed by the essential post-processed outcome on the slice no. $37$ for class level $L=8$ with four distinct activation schemes ($\eta_{\beta}, \eta_{\chi}, \eta_{\xi}, \eta_{\nu}$) are shown in Figure~\ref{fig:C8-37}. It is evident from the experimental data provided in Table~\ref{tab1} that the proposed QFS-Net performs optimally for the $8$-connected quantum fuzzy pixel information heterogeneity assisted activation ($\eta_\xi$) with $L=8$ and gray scale set $S_2$ in comparison with other thresholding schemes and gray scale sets under the four evaluation parameters ($ACC, DS, PPV, SS$)~\cite{alex}. The segmented tumors obtained using the proposed self-supervised procedure under $L=8$ class transition levels with four different thresholding schemes $\eta_\beta$, $\eta_\chi$, $\eta_\xi$ and $\eta_\nu$ are demonstrated in Figures~\ref{fig:C8-S1}-~\ref{fig:C8-S2} for the class boundary sets $S_1$ and $S_2$~\cite{konar4}, respectively. The segmented images using the remaining two class boundary sets ($S_3$ and $S_4$)~\cite{konar4} are provided in the supplementary materials section. The segmented ROIs describing the whole tumor region after the masking procedure using QIS-Net, U-Net and URes-Net are also reported in Figure~\ref{fig:CNN}.

Table~\ref{tab2} presents the numerical results obtained using the proposed QFS-Net and QIS-Net~\cite{konar4} on evaluating the average accuracy ($ACC$), dice similarity score ($DS$), positive prediction value ($PPV$), and sensitivity ($SS$) as reported under $L=8$ class transition levels ($S_1, S_2, S_3, S_4$)~\cite{konar4} with four distinct thresholding schemes ($\eta_\beta$, $\eta_\chi$, $\eta_\xi$ and $\eta_\nu$). The average number of iterations required to converge for each class boundary set is also reported in Table~\ref{tab2}. In addition, Table~\ref{tab3} summarises the results obtained using convolutional U-Net~\cite{olaf} and Residual U-Net (URes-Net)~\cite{guerrero} architectures for two distinct convolutional masks with size $3 \times 3$ and $5\times 5$ with stride sizes of $1$ and $2$. However, the convolutional based architectures (U-Net and URes-Net) marginally outperform our proposed qutrit-inspired fully self-supervised quantum neural network model QFS-Net and the previously developed QIS-Net~\cite{konar4} based on \emph{qubits}. The box plots are also demonstrated in the supplementary materials section citing the outcome reported in Tables~\ref{tab2} and ~\ref{tab3}, respectively. Moreover, to show the effectiveness of our proposed QFS-Net over QIS-Net, U-net and URes-Net, we have conducted one-sided two-sample Kolmogorov-Smirnov (KS)~\cite{gail} test with significance level $\alpha = 0.05$. It is interesting to note that in spite being a fully self-supervised quantum learning inspired by \emph{qutrits}, the QFS-Net has shown similar accuracy ($ACC$) and dice similarity ($DS$) compared with U-Net~\cite{olaf} and URes-Net~\cite{guerrero}. Hence, it can be concluded, that the performance of the QFS-Net model on Dynamic Susceptibility Contrast (DSC) brain MR images is statistically significant and offers a potential alternative to the solution of deep learning technologies. 

\begin{table*}[t]
	\begin{center}
		\caption{Segmented accuracy, dice similarity score, PPV and sensitivity for the slice \#$37$~\cite{data} using QFS-Net}
		\begin{tabular}{p{20pt}p{10pt}p{10pt}p{10pt}p{10pt}p{10pt}p{1pt}p{10pt}p{10pt}p{10pt}p{10pt}p{1pt}p{10pt}p{10pt}p{10pt}p{10pt}p{1pt}p{10pt}p{10pt}p{10pt}p{10pt}}
			\hline	
			\multirow{2}{*}{\centering{\textbf{Level}}} &
			\multirow{2}{*}{\centering{\textbf{Set}}} &
			\multicolumn{4}{p{45pt}}{\centering{$\textbf{ACC} = \frac{T_{RP}+T_{RN}}{T_{RP}+F_{LP}+T_{RN}+F_{LN}}$}} &
			\multicolumn{1}{p{1pt}}{}&
			\multicolumn{4}{p{40pt}}{\centering{$\textbf{DS}=\frac{2T_{RP}}{2T_{RP}+F_{LP}+F_{LN}}$}} &
			\multicolumn{1}{p{1pt}}{} &
			\multicolumn{4}{p{40pt}}{\centering{$\textbf{PPV}=\frac{T_{RP}}{T_{RP}+F_{LP}}$}} &
			\multicolumn{1}{p{1pt}}{} &
			\multicolumn{4}{p{40pt}}{\centering{$\textbf{SS}=\frac{T_{RP}}{T_{RP}+F_{LN}}$}}\\
			\cline{3-6}
			\cline{8-11}
			\cline{13-16}
			\cline{18-21}
			&  & $\eta_\beta$ & $\eta_\chi$ & $\eta_\xi$ & $\eta_\nu$ & & $\eta_\beta$ & $\eta_\chi$ & $\eta_\xi$ & $\eta_\nu$ & & $\eta_\beta$ & $\eta_\chi$ & $\eta_\xi$ & $\eta_\nu$  & & $\eta_\beta$ & $\eta_\chi$ & $\eta_\xi$ & $\eta_\nu$\\
			\hline
			\multirow{4}{*}{$L=4$} & $S_1$ & $\textbf{0.99}$ & $\textbf{0.99}$ & $\textbf{0.99}$ & $\textbf{0.99}$ & & $0.82$ & $0.79$ & $0.79$ & $0.71$ & & $\textbf{0.71}$ & $0.65$ & $0.65$ & $0.65$ & & $0.98$ & $\textbf{0.99}$ & $\textbf{0.99}$ & $\textbf{0.99}$ \\
			& $S_2$ & $\textbf{0.99}$ & $\textbf{0.99}$ & $\textbf{0.99}$ & $\textbf{0.99}$ & & $0.75$ & $\textbf{0.82}$ & $0.78$ & $\textbf{0.82}$ & & $0.71$ & $\textbf{0.72}$ & $0.65$ & $0.71$ & & $0.98$ & $0.98$ & $\emph{0.99}$ & $0.98$ \\
			& $S_3$ & $\textbf{0.99}$ &	$\textbf{0.99}$ & $\textbf{0.99}$ & $\textbf{0.99}$ & & $0.82$ & $\textbf{0.83}$ & $0.78$ & $0.82$ & & $0.71$ & $0.72$ & $0.65$ & $0.71$  & & $0.98$ & $\textbf{0.97}$ & $0.99$ & $0.98$ \\
			& $S_4$ & $\textbf{0.99}$ &	$\textbf{0.99}$ & $\textbf{0.99}$ & $\textbf{0.99}$  & & $0.82$ & $\textbf{0.83}$ & $0.78$ & $0.82$ & & $0.71$ & $0.72$ & $0.65$ & $0.71$ & & $0.98$ & $0.97$ & $\textbf{0.99}$ & $0.98$ \\
			\hline
			\multirow{4}{*}{$L=6$} & $S_1$ & $\textbf{0.99}$ & $\textbf{0.99}$ & $\textbf{0.99}$ & $\textbf{0.99}$ & & $0.82$ & $\textbf{0.83}$ & $0.78$ & $0.82$ & & $0.71$ & $0.72$ & $0.65$ & $0.71$ & & $0.98$ & $0.97$ & $\textbf{0.99}$ & $0.98$ \\
			& $S_2$ & $\textbf{0.99}$ &	$\textbf{0.99}$ & $\textbf{0.99}$ & $\textbf{0.99}$ & & $\textbf{0.82}$ & $0.80$ & $\textbf{0.81}$ & $0.82$ & & $0.71$ & $0.72$ & $\textbf{0.74}$ & $0.71$ & & $0.98$ & $0.89$ & $0.89$ & $0.98$ \\
			& $S_3$ & $\textbf{0.99}$ &	$\textbf{0.99}$ & $\textbf{0.99}$ & $\textbf{0.99}$ & & $0.82$ & $\textbf{0.83}$ & $0.78$ & $0.82$ & & $0.71$ & $\textbf{0.72}$ & $0.65$ & $0.71$ & & $0.98$ & $0.97$ & $\textbf{0.99}$ & $0.98$ \\
			& $S_4$ & $\textbf{0.99}$ &	$\textbf{0.99}$ & $\textbf{0.99}$ & $\textbf{0.99}$ & & $0.82$ & $\textbf{0.83}$ & $0.78$ & $0.82$  & & $\textbf{0.71}$ & $0.72$ & $0.65$ & $0.71$ & & $\textbf{0.98}$ & $0.97$ & $0.99$ & $\textbf{0.98}$ \\
			\hline
			\multirow{4}{*}{$L=8$} & $S_1$ & $\textbf{0.99}$ & $\textbf{0.99}$ & $\textbf{0.99}$ & $\textbf{0.99}$ & & $0.83$ & $0.83$ & $0.82$ & $0.82$ & & $0.74$ & $0.74$ & $0.73$ & $0.74$ & & $0.93$ & $0.93$ & $\textbf{0.94}$ & $0.93$ \\
			& $S_2$ & $\textbf{0.99}$ &	$\textbf{0.99}$ & $\textbf{0.99}$ & $\textbf{0.99}$ & & $\textbf{0.83}$ & $\textbf{0.83}$ & $\textbf{0.83}$ & $\textbf{0.83}$ & & $0.73$ & $0.73$ & $0.73$ & $0.73$ & & $0.95$ & $\textbf{0.96}$ & $\textbf{0.96}$ & $\textbf{0.96}$ \\
			& $S_3$ & $\textbf{0.99}$ &	$\textbf{0.99}$ & $\textbf{0.99}$ & $\textbf{0.99}$ & & $\textbf{0.83}$ & $0.82$ & $0.82$ & $0.82$ & & $\textbf{0.77}$ & $\textbf{0.77}$ & $\textbf{0.77}$ & $\textbf{0.77}$ & & $\textbf{0.89}$ & $\textbf{0.89}$ & $\textbf{0.89}$ & $\textbf{0.89}$ \\
			& $S_4$ & $\textbf{0.99}$ &	$\textbf{0.99}$ & $\textbf{0.99}$ & $\textbf{0.99}$ & & $0.82$ & $0.82$ & $0.82$ & $\textbf{0.83}$ & & $0.73$ & $0.73$ & $0.73$ & $\textbf{0.74}$ & & $\textbf{0.94}$ & $0.94$ & $\textbf{0.94}$ & $\textbf{0.94}$ \\
			\hline
			\label{tab1}
		\end{tabular}
	\end{center}
\end{table*}

\begin{table*}[t]
	\begin{center}
			\caption{Average performance analyses of QFS-Net and QIS-Net~\cite{konar4} for four distinct class levels and activation [One sided non-parametric two sample KS test~\cite{gail} with $\alpha=0.05$ significance level has been conducted and marked in bold.]}
		\begin{tabular}{p{30pt}p{10pt}p{12pt}p{12pt}p{12pt}p{12pt}p{1pt}p{12pt}p{12pt}p{12pt}p{12pt}p{1pt}p{12pt}p{12pt}p{12pt}p{12pt}p{1pt}p{12pt}p{12pt}p{12pt}p{12pt}p{12pt}}
			\hline	
			\multirow{2}{*}{\centering{\textbf{Network}}} &
			\multirow{2}{*}{\centering{\textbf{Set}}} &
			\multicolumn{4}{p{48pt}}{\centering{\textbf{ACC}}} &
			\multicolumn{1}{p{1pt}}{}&
			\multicolumn{4}{p{48pt}}{\centering{\textbf{DS}}} &
			\multicolumn{1}{p{1pt}}{} &
			\multicolumn{4}{p{48pt}}{\centering{\textbf{PPV}}} &
			\multicolumn{1}{p{1pt}}{} &
			\multicolumn{4}{p{48pt}}{\centering{\textbf{SS}}} &
			\multicolumn{1}{p{12pt}}{\centering{\textbf{Avg. $\#Iteration$}}}\\
			\cline{3-6}
			\cline{8-11}
			\cline{13-16}
			\cline{18-21}
			&  & $\eta_\beta$ & $\eta_\chi$ & $\eta_\xi$ & $\eta_\nu$ & & $\eta_\beta$ & $\eta_\chi$ & $\eta_\xi$ & $\eta_\nu$ & & $\eta_\beta$ & $\eta_\chi$ & $\eta_\xi$ & $\eta_\nu$  & & $\eta_\beta$ & $\eta_\chi$ & $\eta_\xi$ & $\eta_\nu$ & \\
			\hline
			\multirow{4}{*}{QFS-Net} & $S_1$ & $\textbf{0.990}$ & $0.987$ & $0.987$ & $0.988$ & & $\textbf{0.799}$ & $0.783$ & $0.782$ & $0.788$ & & $0.713$ & $0.695$ & $0.691$ & $0.698$ & & $0.955$ & $0.954$ & $0.957$ & $0.957$ & $10.78$ \\
			& $S_2$ & $\textbf{0.989}$ & $\textbf{0.989}$ & $0.988$ & $0.987$ & & $\textbf{0.790}$ & $\textbf{0.790}$ & $0.776$ & $0.773$ & & $0.697$ & $0.696$ & $0.679$ & $0.679$ & & $0.957$ & $0.958$ & $\textbf{0.960}$ & $\textbf{0.959}$ & $11.06$\\
			& $S_3$ & $\textbf{0.989}$ & $\textbf{0.989}$ & $\textbf{0.990}$ & $\textbf{0.989}$ & & $0.783$ & $\textbf{0.798}$ & $\textbf{0.795}$ & $0.782$ & & $0.690$ & $0.710$ & $0.718$ & $0.687$ & & $0.955$ & $0.957$ & $0.935$ & $\textbf{0.959}$ & $10.98$\\
			& $S_4$ & $\textbf{0.989}$ & $0.986$ & $0.988$ & $\textbf{0.989}$ & & $0.781$ & $0.767$ & $0.783$ & $\emph{0.800}$ & & $0.694$ & $0.676$ & $0.693$ & $0.713$ & & $0.954$ & $0.955$	& $0.954$ & $0.957$ & $12.12$ \\
			\hline
			\multirow{4}{*}{QIS-Net} & $S_1$ & $0.986$ & $0.987$ & $0.986$ & $0.986$ & & $0.784$  & $0.771$ & $0.767$ & $0.766$ & & $0.698$ & $0.688$ & $0.680$ & $0.672$ & & $0.956$ & $0.947$ & $0.951$ & $\textbf{0.960}$ & $11.77$ \\
			& $S_2$ & $0.987$ & $0.987$ & $0.988$ & $0.988$ & & $0.764$	& $0.761$ & $0.766$	& $0.766$ & & $0.665$	& $0.663$ & $0.667$ & $0.666$ & & $\textbf{0.960}$ & $\textbf{0.959}$ & $\textbf{0.961}$ & $\textbf{0.961}$ & $12.65$\\
			& $S_3$ & $0.986$ & $0.986$ & $0.986$ & $0.987$ & & $0.768$ & $0.781$ & $0.755$ & $0.764$ & & $0.676$ & $0.666$ & $0.659$ & $0.665$ & & $0.955$ & $0.957$ & $0.957$ & $\textbf{0.959}$ & $12.15$\\
			& $S_4$ & $0.987$ & $0.986$ & $0.986$ & $0.986$ & & $0.773$ & $0.764$ & $0.761$ & $0.768$ & & $0.679$ & $0.674$ & $0.668$ & $0.676$ & & $\textbf{0.959}$ & $0.954$ & $0.955$ & $0.957$ & $13.16$\\
			\hline
			\label{tab2}
		\end{tabular}
	\end{center}
\end{table*}
\begin{table}[t]
	\begin{center}
		\caption{Performance analyses of U-Net~\cite{olaf} and URes-Net~\cite{guerrero} for four distinct class levels and activation [One sided non-parametric two sample KS test~\cite{gail} with $\alpha=0.05$ significance level has been conducted and marked in bold.]}
		\begin{tabular}{p{35pt}p{35pt}p{20pt}p{20pt}p{20pt}p{20pt}p{20pt}}
			\hline
			\multirow{1}{*}{\centering{\textbf{Networks}}} & \multirow{1}{*}{\centering{\textbf{Conv-Mask}}} & \multirow{1}{*}{\centering{\textbf{Stride}}} & $\textbf{ACC}$ & $\textbf{DS}$ & $\textbf{PPV}$ & $\textbf{SS}$ \\
			\hline
			\multirow{4}{*}{\textbf{U-Net}} & $3\times 3$ & $1$ & $\textbf{0.993}$ & $\textbf{0.795}$ & $0.717$ & $0.939$ \\ 
			&  $3\times 3$ & $2$  & $\textbf{0.991}$ & $\textbf{0.794}$ & $0.715$	& $0.937$\\	
			&  $5\times 5$ & $1$ & $\textbf{0.996}$ & $\textbf{0.795}$ & $0.726$ & $0.938$ \\
			&  $5\times 5$ & $2$ & $\textbf{0.990}$ & $\textbf{0.797}$ & $0.718$ & $0.940$ \\
			\hline
		    \multirow{4}{*}{\textbf{URes-Net}} & $3\times 3$ & $1$ & $\textbf{0.999}$ & $\textbf{0.806}$ & $0.734$	& $0.932$ \\ 
			&  $3\times 3$ & $2$  & $\textbf{0.997}$ & $\textbf{0.809}$ & $\textbf{0.727}$	& $0.936$\\	
			&  $5\times 5$ & $1$ & $\textbf{0.998}$ & $\textbf{0.805}$ & $\textbf{0.729}$ & $0.939$ \\
			&  $5\times 5$ & $2$ & $\textbf{0.991}$ & $\textbf{0.796}$ & $0.717$ & $0.937$ \\
			\hline
		    \label{tab3}
		\end{tabular}
	\end{center}
\end{table}

\section{Conclusion}
\label{discuss}

An automated brain tumor segmentation using a fully self-supervised QFS-Net encompassing a qutrit-inspired quantum neural network model is presented in this work. The pixel intensities and interconnection weight matrix are expressed in quantum formalism on classical simulations, thereby reducing the computational overhead and enabling faster convergence of the network states. This intrinsic property of the quantum fully self-supervised neural network model allows attaining accurate and time-efficient segmentation in real-time. The suggested QFS-Net achieves high accuracy and dice similarity in spite of being a fully self-supervised neural network model.\\
The proposed quantum neural network model approach is also a faithful mapping towards quantum hardware circuit, and it can also be implemented using quantum gates along with its classical counterparts. The proposed QFS-Net model offers the possibilities of entanglement and superposition in the network architecture, which are often missing in the classical implementations. However, it is also worth noting that the suggested qutrit-inspired fully self-supervised quantum neural network model is computed and experimented on a classical system. Hence, the proposed model architecture is not quantum in a real sense, instead it is quantum-inspired. It is also worth noting that the QFS-Net is validated solely for complete tumor and the network has potential for multi-level segmentation which is evident from the segmented brain MR lesions. Nevertheless, it remains an uphill task to optimize the hyper-parameters for obtaining optimal multi-class segmentation. Authors are currently engaged in this direction.




\appendix
\subsection{Convergence analysis of QFS-Net}
\label{conv:QFS-Net}

Let us consider the optimal phase angles for the weighted matrix and the activation are denoted as $\overline \omega$ and $\overline \gamma$, respectively and defined as follows:
\begin{equation}
\upsilon^{\iota}=\omega^{\iota}-  \overline \omega
\end{equation}
\begin{equation}
\mu^{\iota}=\gamma^{\iota}- \overline \gamma
\end{equation}
and
\begin{equation}
\delta^{\iota}=\omega^{\iota+1}- \omega^{\iota}=\upsilon^{\iota+1}-\upsilon^{\iota}
\end{equation}
\begin{equation}
\rho^{\iota}=\gamma^{\iota+1}- \gamma^{\iota}=\mu^{\iota+1}-\mu^{\iota}
\end{equation}
Also, the derivative of the loss function $\zeta (\omega,\gamma)$ with respect to $\omega,\gamma$ is depicted as follows.
\begin{equation}
\begin{split}
\frac{\partial \zeta(\omega,\gamma)}{\partial \omega_{ik}} = \frac{2}{N} \sum_{i}^N \sum_{k=1}^8 \triangle \Theta_{ik}(\omega_{ik},\gamma_{ik})^{\iota}\\\left[\frac{\partial \Theta_{ik}(\omega_{ik},\gamma_{ik})^{\iota+1}}{\partial \omega_{ik}} - \frac{\partial \Theta_{ik}(\omega_{ik},\gamma_{ik})^{\iota}}{\partial \omega_{ik}}\right]
\end{split}
\end{equation}
\begin{equation}
\begin{split}
\frac{\partial \zeta(\omega,\gamma)}{\partial \gamma_i} = \\\frac{2}{N}\sum_{i}^N \triangle \Theta_{i} (\omega_{i},\gamma_i)^{\iota}\left[\frac{\partial \Theta_{i}(\omega_{i},\gamma_i)^{\iota+1}}{\partial \gamma_i} - \frac{\partial \Theta_{i}(\omega_{i},\gamma_i)^{\iota}}{\partial \gamma_i}\right]
\end{split}
\end{equation}
where
\begin{equation}
\triangle \Theta_{ik} (\omega_i,\gamma_{ik})^{\iota} = |\Theta_{ik}(\omega_{ik},\gamma_i) ^{\iota+1}-\Theta_{ik}(\omega_{ik}, \gamma_i)^{\iota}|
\end{equation}
and
\begin{equation}
\begin{split}
\Theta_{ik}(\omega_{ik},\gamma_i)^{\iota} =\left[Im(\mathcal{H}\{\langle \theta_{ik}^{\iota}|\xi_i^{\iota}\rangle\})\right]^2 = \\\left[Im(\cos(\omega_{ik}-\gamma_i)^{\iota} +\jmath \sin(\omega_{ik}-\gamma_i)^{\iota})\right]^2
\end{split}
\end{equation}
The change in phase or angles ($\triangle \omega$ and $\triangle \gamma$) in the Hadamard gate are evaluated using the following equations.
\begin{equation}
\triangle \omega_{ik}^{\iota}=-\sigma_{ik} \{\frac{\partial \zeta(\omega,\gamma)^{\iota}}{\partial \omega_{ik}^{\iota}}\zeta (\omega,\gamma)^{\iota}\}^{\frac{1}{t}}
\end{equation}
\begin{equation}
\triangle \gamma_i^{\iota}=-\sigma_{i} \{\frac{\partial \zeta(\omega,\gamma)}{\partial \gamma_i^{\iota}} \zeta (\omega,\gamma)^{\iota}\}^{\frac{1}{t}}
\end{equation}
where, the learning rate for the self-supervised updating of the weights in QFS-Net is denoted as $\sigma_{ik}$. It is computed using the relative difference between the candidate and its neighborhood \emph{qutrit} neurons (intensities) with  $t > 2$ as
\begin{equation}
\sigma_{ik} = \mu_i-\mu_{ik} \forall k =1,2 \dots 8 
\end{equation}
Similarly, the learning rate for updating the activation is denoted as $\sigma_i$ and is equal to the quantum fuzzy contribution of the candidate neuron ($\mu_i$). 
The conditions for the super-linear convergence of the sequences of $\{\omega^{\iota}\}$ and $\{\gamma^{\iota}\}$ can be formulated as~\cite{zhen} 
\begin{equation}
\lim_{\iota \rightarrow \infty} \frac{||\omega^{\iota+1}-\overline \omega||}{||\omega^{\iota}-\overline \omega||}\leq 1  
\label{lim1}
\end{equation}
and
\begin{equation}
||\upsilon^{\iota+1}||= O||\delta^{\iota}||
\end{equation}
Also,
\begin{equation}
\lim_{\iota \rightarrow \infty} \frac{||\gamma^{\iota+1}-\overline \gamma||}{||\gamma^{\iota}-\overline \gamma||}\leq 1
\label{lim2}
\end{equation}
and 
\begin{equation}
||\mu^{\iota+1}||=O||\rho^{\iota}||
\end{equation}
In order to prove the convergence of the sequences of $\{\omega^{\iota}\}$ and $\{\gamma^{\iota}\}$, according to Thaler theorem, we obtain
\begin{eqnarray}
  \zeta (\omega^{\iota+1}, \gamma^{\iota+1})-\zeta (\omega^{\iota}, \gamma^{\iota}) =  \\ \nonumber
  \left[
    \begin{array}{cc}
      \triangle \omega_{ik}^{\iota} & \triangle \gamma_{i}^{\iota} \\
    \end{array}
  \right]\left[
           \begin{array}{c}
             \frac{\partial \zeta(\omega,\gamma)^{\iota}}{\partial \omega_{ik}^{\iota}} \\
             \frac{\partial \zeta(\omega,\gamma)^{\iota}}{\partial \gamma_{ik}^{\iota}} \\
           \end{array}
         \right]+O \left[
    \begin{array}{cc}
      ||\triangle \omega_{ik}^{\iota} & \triangle \gamma_{i}^{\iota}|| \\
    \end{array}
  \right]
\end{eqnarray}
\begin{equation}
\approx \left[\{-\sigma_{ik}\frac{\partial \zeta(\omega,\gamma)^{\iota}}{\partial \omega_{ik}^{\iota}}\}^2+
\{-\sigma_i\frac{\partial \zeta(\omega,\gamma)^{\iota}}{\partial \gamma_{ik}^{\iota}}\}^2 \right] \{\zeta (\omega^{\iota}, \gamma^{\iota})\}^{\frac{1}{t}}
\end{equation}
Hence, ($\zeta (\omega^{\iota+1}, \gamma^{\iota+1})-\zeta (\omega^{\iota}, \gamma^{\iota}))\leq 0$ and it is clearly evident that the sequences of $\{\omega^{\iota}\}$ and $\{\gamma^{\iota}\}$ are monotonically decreasing. The coherent nature of these two sequence leads to the following. 
\begin{equation}
\lim_{\iota \rightarrow \infty} \zeta (\omega^{\iota}, \gamma^{\iota})=(\overline \omega, \overline \gamma)
\label{lim3}
\end{equation}
The rapid convergence of the iteration sequences $\{\omega^{\iota}\}$ and $\{\gamma^{\iota}\}$ are due to
\begin{equation}
\lim_{\iota \rightarrow \infty} \frac{||\zeta (\omega^{\iota+1}, \gamma^{\iota+1})-(\overline \omega, \overline \gamma)||}{||\zeta (\omega^{\iota}, \gamma^{\iota})-(\overline \omega, \overline \gamma)||} \leq 1
\end{equation}
The super-linear convergence of the sequences can be shown as follows. \\
Let $G_{\omega}=\frac{\partial \zeta(\omega,\gamma)^{\iota}}{\partial \omega_{ik}^{\iota}}$, then
\begin{equation}
\frac{||\omega^{\iota +1}||}{||\delta^{\iota}||}=\frac{||\omega^{\iota+1}-\overline \omega||}{||-\sigma_{ik} \{\frac{\partial \zeta(\omega,\gamma)^{\iota}}{\partial \omega_{ik}^{\iota}}\zeta (\omega,\gamma)^{\iota}\}^{\frac{1}{t}}||} \geq \frac{||\omega^{\iota+1}-\overline \omega||}{\sigma_{ik} G_{\omega} \{\zeta (\omega,\gamma)^{\iota}\}^{\frac{1}{t}}}
\end{equation}
Hence, 
\begin{equation}
||\omega^{\iota+1}-\overline \omega||=O(\{\zeta (\omega,\gamma)^{\iota}\}^{\frac{1}{t}}\})
\end{equation}
Consequently,
\begin{equation}
||\omega^{\iota +1}|| = O(||\delta^{\iota}||)
\end{equation}
which proves that the convergence behavior of the iteration sequence $\{\omega^{\iota}\}$ is super-linearly convergent. \\
Similarly, let $G_{\gamma}=\frac{\partial \zeta(\omega,\gamma)^{\iota}}{\partial \gamma_{ik}^{\iota}}$, then
\begin{equation}
\frac{||\gamma^{\iota +1}||}{||\rho^{\iota}||}=\frac{||\gamma^{\iota+1}-\overline \gamma||}{||-\sigma_i \{\frac{\partial \zeta(\omega,\gamma)^{\iota}}{\partial \gamma_i^{\iota}}\zeta (\omega,\gamma)^{\iota}\}^{\frac{1}{t}}||} \geq \frac{||\gamma^{\iota+1}-\overline \gamma||||}{\sigma_i G_{\gamma} \{\zeta (\omega,\gamma)^{\iota}\}^{\frac{1}{t}}}
\end{equation}
Hence, 
\begin{equation}
||\gamma^{\iota+1}-\overline \gamma||=O(\{\zeta (\omega,\gamma)^{\iota}\}^{\frac{1}{t}}\})
\end{equation}
Consequently,
\begin{equation}
||\gamma^{\iota +1}|| = O(||\rho^{\iota}||)
\end{equation}
which proves that the convergence behavior of the iteration sequence $\{\gamma^{\iota}\}$ is super-linearly convergent.

\end{document}